\documentclass[11pt]{article}

\usepackage{color}
\usepackage{latexsym}
\usepackage{amssymb}
\usepackage{epsf}
\usepackage{amsmath}
\usepackage{graphicx}
\usepackage{slashed}
\usepackage{float}
\usepackage[square,comma,numbers,sort&compress]{natbib}
\DeclareMathOperator{\BR}{BR}

\newcommand{\al}[1]{\begin{align}#1\end{align}}
\newcommand{\paren}[1]{\left(#1\right)}
\newcommand{\sqbr}[1]{\left[#1\right]}

\newcommand{\fn}[1]{\!\left(#1\right)}
\newcommand{\br}[1]{\left\{#1\right\}}
\newcommand{\nn}{\nonumber\\}
\newcommand{\bb}{\begin{bmatrix}}
\newcommand{\eb}{\end{bmatrix}}

\newcommand{\MKK}{M_\text{KK}}

\newcommand{\Q}{\mathcal{Q}}

\newcommand{\te}{\text}

\usepackage{a4wide}
\allowdisplaybreaks[2]

\begin{document}



\title{\huge 
{Universal extra dimensions after Higgs discovery}
\bigskip}

\author{\Large
Takuya Kakuda,\footnote{
E-mail: {kakuda@muse.sc.niigata-u.ac.jp}
}
\ 
Kenji Nishiwaki,\footnote{
E-mail: {nishiwaki@hri.res.in}
}
\ 
Kin-ya Oda,\footnote{
E-mail: {odakin@phys.sci.osaka-u.ac.jp}
}\smallskip\\
\Large
and Ryoutaro Watanabe\footnote{
E-mail: {wryou@post.kek.jp}
}
\bigskip\\
\it $^*$ Department of Physics, Niigata University, Niigata 950-2181, Japan\smallskip\\
\it $^\dagger$ Regional Centre for Accelerator-based Particle Physics\smallskip\\
\it Harish-Chandra Research Institute, Allahabad 211 019, India\smallskip\\
\it $^\ddag$ Department of Physics, Osaka University, Osaka 560-0043, Japan\smallskip\\
\it $^\S$ Theory Group, KEK, Tsukuba, Ibaraki 305-0801, JAPAN\smallskip\\
\bigskip}





\maketitle

\begin{abstract}
\normalsize
\noindent
We show bounds on {five- and six-dimensional} universal extra dimension (UED) models from the latest results of the Higgs searches at the LHC and {from} the electroweak precision data for the $S$ and $T$ parameters. 
{We consider} the minimal UED model in five dimensions and {the ones in six dimensions}, {compactified} on $T^2/Z_2$, $T^2/(Z_2 \times Z_2')$, $ T^2/Z_4$, $S^2$, $S^2/Z_2$, {the real projective plane,} and the projective sphere. 
The highest possible ultraviolet cutoff scale for each UED model is evaluated from the {electroweak} vacuum stability by solving the renormalization group equation of the Higgs {self-coupling}.
{This scale turns out to be lower than the conventional one obtained from the perturbativity of the gauge coupling.}
{The resultant 95\% {C.L.} lower bounds on the first {Kaluza--Klein} scale from the LHC results and from the $S, T$ analysis are {600} and 700\,GeV in the minimal UED model, while those in the {six-dimensional} UED models are {800}--${1300}\,\text{GeV}$ and $900$--$1500\,\text{GeV}$, respectively.}
\end{abstract}
\vfill
\begin{flushright}
HRI-P-13-02-002\\
RECAPP-HRI-2013-003\\
KEK-TH-1631\\
OU-HET-789
\end{flushright}

\newpage




\section{Introduction}
The ATLAS and CMS experiments at the CERN {LHC} have observed a particle around 126\,GeV{,} which is consistent with the Standard Model (SM) Higgs boson~\cite{:2012gk,:2012gu}.
The signal strength (defined as the ratio of {the production} cross section {times the branching ratio of the observed particle} {to} {that of} the SM Higgs)
for {the} decay {channels} into diphoton ($\gamma\gamma$) and diboson ($ZZ$ and $WW$) {are reported in Refs.}~\cite{ATLAS:2013gamma,ATLAS:2013Z,ATLAS:2013W,CMS:2013gamma,CMS:2013Z,CMS:2013W}. 
Namely, the signal strengths of $H \to \gamma\gamma$, $ZZ${,} and $WW$ {have} turn{ed} out to be $1.65\pm0.24^{+0.25}_{-0.18}$, $1.7^{+0.5}_{-0.4}${,} and $1.01\pm0.31$ at the ATLAS experiment, 
{and} $0.78\pm0.27$ ({multivariate analysis} based), $0.91^{+0.30}_{-0.24}${,} and $0.71\pm0.37$ (cut based) at the CMS experiment{, respectively}. 
{All} these results are consistent with the SM {within less than 2$\sigma$}{,} but there still {remains} {room} for a new physics effect.
See{,} e.g.{,}\ Refs.~\cite{Azatov:2012bz,LHCHiggsCrossSectionWorkingGroup:2012nn,Moreau:2012da,Cacciapaglia:2012wb,Falkowski:2013dza,Giardino:2013bma,Alanne:2013dra,Ellis:2013lra,Djouadi:2013qya} for analyses based on effective Lagrangian methods.\footnote{
{Global fits of the minimal UED scenario {were} done in Ref.~\cite{Bertone:2010ww}}.
}
{In this paper, we study the constraints} on
the {existence of the} universal extra dimensions ({UEDs})~\cite{Appelquist:2000nn}\footnote{
See also Ref.~\cite{Antoniadis:1990ew} for an earlier proposal of a TeV scale extra dimension.
}
{from the Higgs searches}.

{In the UED} scenario, {each} SM field propagates in {one or more} compactified extra dimensions and {is accompanied by} {its} massive copies, {called} {Kaluza--Klein} (KK) particles.
{Already} in the simplest five-dimensional (5D) minimal UED (mUED) {model} on the {orbifold} $S^1/Z_2$~\cite{Appelquist:2000nn}, in which no {tree-level} brane-localized term {is assumed at an {UV} cutoff scale of the 5D gauge theory}, {there} {exists} {an} attractive {feature}:
The lightest KK particles {become} stable due to the symmetry in the geometry, the KK parity, and serves as a natural dark matter candidate~\cite{Servant:2002aq,Belanger:2010yx}.
The KK particles are expected to {exist above} around $1\,\text{TeV}$, {which is} consistent {to} the indirect {bound from} the {$S$, $T$} parameters, $M_{\text{KK}} \gtrsim 700\,\text{GeV}$~\cite{Baak:2011ze}, and {from} the $b \to s \gamma$ process, $M_{\text{KK}} \gtrsim 600\,\text{GeV}$~\cite{Haisch:2007vb}, {with} $M_{\text{KK}}$ 
{being the first KK mass.}
The prospects of the mUED at the LHC and future linear colliders have been discussed {in Refs.}~\cite{Rizzo:2001sd,Macesanu:2002db,Cheng:2002ab,Carone:2003ms,Bhattacharyya:2005vm,Bhattacherjee:2005qe,Cembranos:2006gt,Bhattacherjee:2007wy,Bhattacherjee:2008ik,Konar:2009ae,Matsumoto:2009tb,Bhattacharyya:2009br,Bandyopadhyay:2009gd,Choudhury:2009kz,Bhattacherjee:2010vm,Murayama:2011hj,Datta:2011vg,Chang:2012wp,Belyaev:2012ai,Edelhauser:2013lia} and {those} in the context of discrimination from other models with similar final states {in Refs.}~\cite{Battaglia:2005zf,Smillie:2005ar,Datta:2005zs,Datta:2005vx,Bhattacherjee:2009jh,Ghosh:2010tp}.

{A} way of {extending} the minimal scenario is {to} consider the model in two-dimensional extra space.
{Models} have been proposed on two-torus, $T^2/Z_2$~\cite{Appelquist:2000nn}, $T^2/Z_4$ ({chiral square})~\cite{Dobrescu:2004zi,Burdman:2005sr}, $T^2/(Z_2 \times Z'_2)$~\cite{Mohapatra:2002ug}, on two-sphere $S^2/Z_2$~\cite{Maru:2009wu}{,} {on} $S^2$ with {a} Stueckhelbarg field{~\cite{Nishiwaki:2011gk,Nishiwaki:2011gm}}, and on the {nonorientable} manifolds: the real projective plane $RP^2$~\cite{Cacciapaglia:2009pa} and the projective sphere (PS)~\cite{Dohi:2010vc}.
{An} advantage {of such a} six-dimensional (6D) {UED} model is that the number of generation is {predicted to be} (a multiple of) three~\cite{Dobrescu:2001ae}, {from the requirement of} the cancellation of {the} 6D gravitational and $SU(2)_L$ global anomalies, which cannot be eliminated via {the} {Green--Schwarz} mechanism.
We can find works on collider phenomenology in the cases of $T^2/Z_4$~\cite{Burdman:2006gy,Dobrescu:2007xf,Ghosh:2008ix,Ghosh:2008dp,Ghosh:2008ji,Choudhury:2011jk} and {of} $RP^2$~\cite{Cacciapaglia:2011hx,Cacciapaglia:2011kz,Cacciapaglia:2012dy,Cacciapaglia:2013wha}.
Recently, other possibilities of generalization of these models {by an introduction of the bulk mass term and/or the brane-localized Lagrangians} have been studied in Refs.~\cite{Bhattacherjee:2008kx,Haba:2009uu,Park:2009cs,Chen:2009gz,Kong:2010xk,Huang:2012kz,Rizzo:2012rb,Flacke:2008ne,Datta:2012xy,Datta:2012tv,Flacke:2012ke,Majee:2013se,Flacke:2013pla}.

In general, the single Higgs production {process} via the loop-induced gluon fusion is enhanced {in a UED model, while} the branching ratio of the Higgs to diphoton is suppressed because of interference effect between bosons and fermions inside the loops. {These UED effects have been first shown in the 5D mUED~\cite{Petriello:2002uu}. In 6D UED, the enhancement of the gluon fusion is studied in Ref.~\cite{Maru:2009cu}{,} and the diphoton decay rate is obtained in Ref.~\cite{Nishiwaki:2011vi}. See Refs.~\cite{Rai:2005vy,Nishiwaki:2011gk,Nishiwaki:2011gm,Belanger:2012mc,Ujjal:2013} for more works {in} this direction.}
In this paper, we {perform} more {elaborated} analysis compared with our previous work in {Ref.}~\cite{Nishiwaki:2011vi}, with {a} varying portion of production channels for each event category.
We also estimate constraints from the {$S$ and $T$} parameters in every model for completeness.
In addition to the effects of the KK Higgs boson and the KK top quark~\cite{Appelquist:2002wb,Baak:2011ze}, {those of} the KK gauge boson {are} taken into account {for the first time in the literature}.

An important {number} in {the} UED phenomenology is the {highest possible} {UV cutoff scale,} $\Lambda_{\text{max}}$, {allowed by the electroweak vacuum stability}.
The cutoff scale $\Lambda$ of a UED model {gives} the upper bound of {the} KK summation in loop processes.
Therefore{,} different values of $\Lambda$ result in different bounds on $M_\text{KK}$.
{In the mUED with the $126\,\text{GeV}$ Higgs},
the highest {possible $\Lambda$} becomes quite low, $\Lambda\lesssim 5 M_{\text{KK}}$~\cite{Bhattacharyya:2002nc}.
{In this paper, we examine} all the 5D and 6D UED models without {resorting to the} approximation {employed in} the analysis of gauge coupling running in our previous work~\cite{Nishiwaki:2011gk}. 
When we consider a model with a low cutoff scale, threshold corrections via {higher-dimensional} operators {can become} significant, {which we will take into account {for the $S$ and $T$} parameter constraints.}
{
Effects from such {higher-dimensional} operators on Higgs signals have already been discussed in Ref.~\cite{Nishiwaki:2011vi}.
}

The paper is organized as follows.
In {Sec.}~\ref{section:2}, we estimate the highest cutoff {scale} of all the UED models.
Based on the results, we calculate {the} direct and indirect bounds {from} the LHC results in {Sec.}~\ref{section:LHC} and {from} the {$S$, $T$} parameters in {Sec.}~\ref{section:ST}. 
In {Sec.}~\ref{section:summary}, we summarize our results and discuss future prospects.
{Detailed} {formulas that} we use in this paper {can be} found in {the appendix}. 


\section{Vacuum stability bound 
\label{section:2}}
In this section, we estimate {the} UV cutoff scale in seven types of six-dimensional UED models {on}
$T^2/Z_2$, {$T^2/(Z_2 \times Z_2')$}, $T^2/Z_4$, $RP^2$; 
$S^2$ with {a} Stueckhelbarg field, $S^2/Z_2${;} and {the} {PS} {and in the 5D mUED}.
The orbifolding in $T^2/Z_2$, {$T^2/(Z_2 \times Z_2')$} {and} $T^2/Z_4$ on {two}-torus ($T^2$) {projects out a} chiral zero mode from each matter fermion{.}
On {two}-sphere $S^2$, we can obtain {a} Weyl {fermion} {in each} zero {mode,} {due to the} {monopolelike} classical configuration {of an extra $U(1)_X$ gauge boson.} 
{However,} we {must} eliminate the {phenomenologically unacceptable} massless zero mode of the $U(1)_X$ gauge boson, and we {will discuss} three possibilities for treating this issue in this paper.

The geometries of $RP^2$ and {PS} are unoriented {and} have no local fixed points. Consequently, their KK mass spectra take unique forms, 
{for which the} pattern {is distinctive} from those of the other UED models.
A brief review of {the} models {studied in this paper} {can also be} found in our previous paper~\cite{Nishiwaki:2011gm}.

To find {the highest possible} UV cutoff {$\Lambda_\text{max}$}, we {examine the} vacuum stability bound by solving {the} {renormalization group equation} (RGE).
{As said above, the} UV cutoff plays an important role {in the estimation of the} KK loop effects in processes {involving} loop diagrams.
In later sections, we {use} the results of this section for calculating the deviations in the single Higgs production processes and the {Peskin--Takeuchi} $S$ and $T$ parameters.
It is noted that{,} in this paper, we {mainly consider the situation where} {radii} of compactified fifth and sixth directions $R_5,\,R_6$ are the same: $R_5=R_6=R$.\footnote{
In the $T^2/Z_4$ model, the condition $R_5=R_6=R$ is automatically realized due to the property of the orbifolding.
{See also {Ref.}~\cite{Lim:2009pj} for a realization of {$CP$} violation from the complex structure of $T^2/Z_4$, which appears in {four-dimensional} effective interactions after KK decomposition.}
}

\subsection{RGE in 6D UED models\label{Section:2-1}}
Considering {the} RGE is an effective way {of} probing scale dependence. 
Its concrete form is derived from {the} invariance, {under the change} of the renormalization scale $\mu$,  of {the} bare vertex function~$\Gamma_0${,} which is a function of bare parameters. 
The scale invariance requires that
\al{
\mu \frac{d}{d \mu}  \Gamma_0 (\{c_{0}\}, \{m_{0}\}, \{\Phi_{0}\}) = 0 \label{2-1},
} 
where $\{c_{0}\}$, $\{m_{0}\}$, and $\{\Phi_0\}$ represent sets of bare couplings, masses, and fields, respectively. 
Since bare parameters and fields are divergent themselves, we can rewrite {the} bare ones with finite physical ones (renormalized parameters and fields) and counter terms, which contain divergences.
In this paper, we show all the bare/renormalized variables with/without the subscript {``$0.$"}

In the following, we consider the RGE for the Higgs quartic coupling $\lambda$ in the 6D models.
{We} obey the convention {of} Ref.~\cite{Denner:1991kt} {in} describing the {electroweak (EW)} sector.
The potential of the Higgs field $H$ at the tree level is depicted as
\al{
{-\frac{M^2_{H0}}{2} H^2_{0} + \frac{\lambda_{(6)0}}{4} H^4_{0},}
}
where $M_{H0}$ and $\lambda_{(6)0}$ are the bare Higgs mass and 6D Higgs couplings.
After {the} 6D bare Higgs field $H_0$ is {KK expanded}, we can find the zero mode $H_0^{(0)}$, where we {use a superscript for a KK index.}
{In considering the one-loop running of $\lambda$,} we need not consider the renormalization of the Higgs mass and hence the physical Higgs mass $m_H$ {becomes}
\al{
m_H = \sqrt{\lambda_{(6)0}} v_{(6)0} = \sqrt{\lambda_{(6)}} v_{(6)} = \sqrt{\lambda} v,
}
where the {four-dimensional (4D) Higgs vacuum expectation value} $v= 246\,\text{GeV}$ and {quartic} coupling {$\lambda$} are expressed as $v_{(6)} = v/\sqrt{V},\ \lambda_{(6)} = \lambda \sqrt{V}$, {with $V$ being the volume of the extra dimensions.}
{Let us write}
\al{
{H_0^{(0)}       = \sqrt{Z_H} H^{(0)}},\quad
\lambda_{0} = Z_{\lambda} (Z_H)^{-2} \lambda,
}
where $Z_{\lambda}$ is the renormalization {factor for} the Higgs quartic coupling and $\sqrt{Z_{H}}$ is that for {the} wave function renormalization of the Higgs {zero mode}.
We also need the information {of} the RGEs for the gauge and Yukawa couplings 
{to compute the running of $\lambda$.}
We summarize the beta functions{,}
\al{
\mu \frac{d}{d\mu} \mathcal{Q} = \beta_{\mathcal{Q}},
}
{where detailed {form of} $\beta_{\mathcal{Q}}$ {can be} found} in Appendix~\ref{Appendix_A}.

{Let us review how to compute} RGEs in a theory with {(a)} compactified extra dimension(s).
We adopt the bottom-up approach discussed in Refs.~\cite{Dienes:1998vh,Dienes:1998vg}, 
which {takes into account} a contribution of a massive particle {to the beta functions} {when the increasing scale $\mu$ passes its mass.}
{In the case of the UED,} after KK decomposition, the corresponding 4D effective theory {contains} not only the SM fields {but} also their {KK partners}.
{Following this prescription, we get}
\al{\label{Eq:betafunction}
{
\beta_\Q = \beta_\Q^{\text{(SM)}} + \sum_{s:\,\text{massive states}} \theta(\mu - M_s) \Big( N_s
\beta^{\text{(NP)}}_{s,\Q} \Big),
}
}
where $\beta_\Q^{\text{(SM)}}$ and {$\beta_{s,\Q}^{\text{(NP)}}$} {are} the contributions from the SM particles and {from the} {new massive} {ones} with mass $M_s$, {respectively,}
and $N_s$ {is} the number of degenerated states. 
At the tree level, $M_s$ is expressed as
\al{
{
M_s^2 = m_{s,(\te{SM})}^2 + M_{s,(\te{KK})}^2,
}
}
{where $m_{s,(\te{SM})}$ is the SM mass {of the corresponding zero mode} and $M_{s,(\te{KK})}$ {are} KK {masses.} 
{In general,} the value of $M_{s,(\te{KK})}$ is much greater than that of $m_{s,(\te{SM})}$
and $M_s^2$ can be approximated {as} $M_s^2 \simeq M_{s,(\te{KK})}^2$.
}


Let us review the KK expansions in the 6D UED models.
In the {models on the orbifolded} $T^2$ {[{namely} $T^2/Z_2$, {$T^2/(Z_2 \times Z_2')$}, {and} $T^2/Z_4$]} and the  {one} on $RP^2$,
{the KK mass} $M_{s,(\te{KK})}^2$ {becomes of $T^2$ type:}
\al{
M_{s,(\te{KK})}^2 \rightarrow M_{(m,n)}^2 := {\frac{m^2}{R_{5}^{2}} + \frac{n^2}{R_{6}^2}}, 
}
where $m$ $(n)$ is the KK index along {the fifth (sixth)} direction {and $N_{(m,n)}=1$ irrespective of $m$ and $n$.}
We {note} that {the} beta functions for gauge, Yukawa{, and Higgs self-couplings} take the same forms {irrespective of} the models {based on $T^2$} and are independent of the KK indices.
This reason is as follows.
Because of the flat profile of the zero modes, the {three-point} functions with one SM field and the {four-point} functions with two SM fields 
{become} universal at their leading order after using the orthonormality of KK mode functions.
{In contrast,} the value $N_s$ and the summation of the KK index $\sum_s$ in Eq.~(\ref{Eq:betafunction}) are affected by the difference in the patterns of the orbifolding. 
Hence{,} {the evolution of $\lambda$} {depends on the choice of the model.}
{The explicit range of $m,n$ {summation} {is} {shown} in Table~\ref{range_of_mandn}.}

\begin{table}[t]
\begin{center}
\begin{tabular}{|c|c|}
\hline 
{Type} of orbifolding & {Range} of $(m,n)$ \\
\hline 
$T^2/Z_2$ & {$m+n \geq 1,$ or $m=-n \geq 1$} \\
$T^2/(Z_2 \times Z'_2)$ & $0 \leq m < \infty, \ 0 \leq n < \infty;\ (m,n) \not=(0,0)$ \\
$T^2/Z_4$ & $1 \leq m < \infty, \ 0 \leq n < \infty$ \\
\hline
\end{tabular}
\caption{The range of the parameters $(m,n)$ except for the zero mode $(m,n)=(0,0)$
in each case of the orbifolding.
\label{range_of_mandn}}
\end{center}
\end{table}%

\begin{figure}[t]
\begin{center}
   \includegraphics[width=0.7\columnwidth]{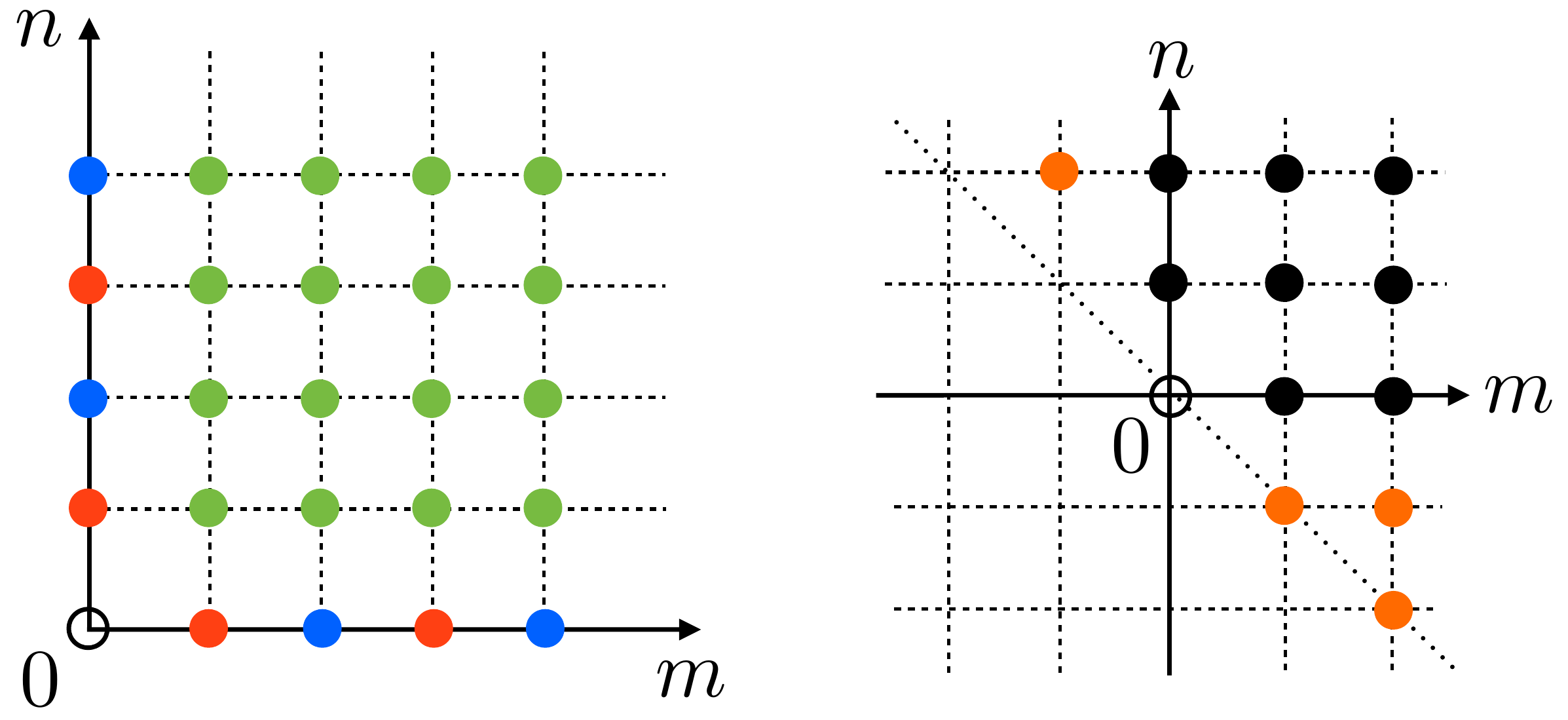}
\caption{{\it Left}\,: The patterns of the remaining bosonic {modes} in {the} $RP^2$ model, where blue, red, and green {points indicate} that {they} belong to the regions I, II, and III, respectively. The {definitions} of {these} regions {are} found in the text.
{\it Right}\,: {The same as left one for {fermonic} modes.}
{At} the {orange} points (region IV), there {are} no bosonic {modes}.
In both pictures, the black circles $(m=n=0)$ correspond to the SM particles.
}
\label{fig:RP2scheme}
\end{center}
\end{figure}

Let us turn to the model on $RP^2$.
{In Fig.~\ref{fig:RP2scheme}, we show the surviving modes.}
The surviving modes of KK fermions {become} the same as in the $T^2/Z_2$ {model}. 
{On the other hand,} the patterns of the bosonic particles are complicated.
The allowed range of $m$ and $n$ {is} $m \geq 0,\ n \geq 0$, {and the type} of surviving mode {is} classified {into} the following {four.} 
In region I{,} $(m,n)=(0,2), (0,4), (0,6), \dots$ and $(m,n)=(2,0), (4,0), (6,0), \dots${; a physical scaler} mode coming from {the} extra {component} of {the} 6D gauge boson is {projected out.} 
In region II{,} $(m,n)=(0,1), (0,3), (0,5), \dots$ and $(m,n)=(1,0), (3,0), (5,0), \dots${;} the only surviving bosonic mode is this scalar {that was projected in {region I}}.
In region III{;} $m \geq 1,\ n \geq 1${;} all the bosonic modes are {left as is, just} like {in other} orbifolded models on $T^2$.
{In the last region IV, only fermionic degrees of freedom remain.}

{Next, we go on to} the {models} based on $S^2$.
The explicit form of the KK mass $M_{s,(\te{KK})}^2$ on $S^2$ is 
\al{
M_{s,(\te{KK})}^2 \rightarrow M_{(j)}^2 := \frac{j(j+1)}{R^2}, 
}
with the index $j \geq 1$.
For each $j$th mode in the $S^2$, $S^2/Z_2$, and PS models, respectively, {the} number of degrees of freedom {reads}
\al{
n^{S^2}(j) = 2j+1,
\label{S2count}
}
\al{
n^{S^2/Z_2}(j) =\begin{cases}j+1 & \quad \text{for }j=\text{{even}}  \\ j & \quad \text{for }j=\text{{odd}}\end{cases},
\label{S2Z2count}
}
\al{
 n_{\text{fermion}}^{\text{PS}} = 2j+1, \quad
 n_\text{even}^\text{PS}(j) = \begin{cases} 2j+1 \\ 0 \end{cases}\!\!\!\!\!\!, \quad
 n_\text{odd}^\text{PS}(j) = \begin{cases} 0 & \quad\text{for }j=\text{{even}} \\ 2j+1 & \quad\text{for }j=\text{{odd}}\end{cases}.
\label{PScount}
}
In the cases of $S^2$ and $S^2/Z_2$, {the} number of the surviving degrees of freedom {is} {the same for} KK bosons and fermions. 
{On the other hand,} {the} {PS} is {similar} to $RP^2${;} {that is,} surviving KK bosons are divided into two categories, even and odd.
The even category includes all the KK {bosons} except {for} the physical scalar from {the} 6D gauge boson, while the odd one only contains this {one.}
We {note} that the number of {degenerate} states {is} $2j+1${,} irrespective {of} the statistics of the particles and their oddness/evenness.
Finally, we comment on the beta functions of the $S^2$-based models. 
{From the surviving} bosonic particles in each KK level, {we can see} that 
the {RGEs} in $S^2$, $S^2/Z_2$ are similar to those in $T^2/Z_2$, {$T^2/(Z_2 \times Z_2')$}, $T^2/Z_4$, {while those in {the PS are similar} to those in $RP^2$.}

\subsection{Running of Higgs {self-coupling} and Vacuum Stability}

\begin{figure}[t]
\begin{center}
   \includegraphics[width=0.99\columnwidth]{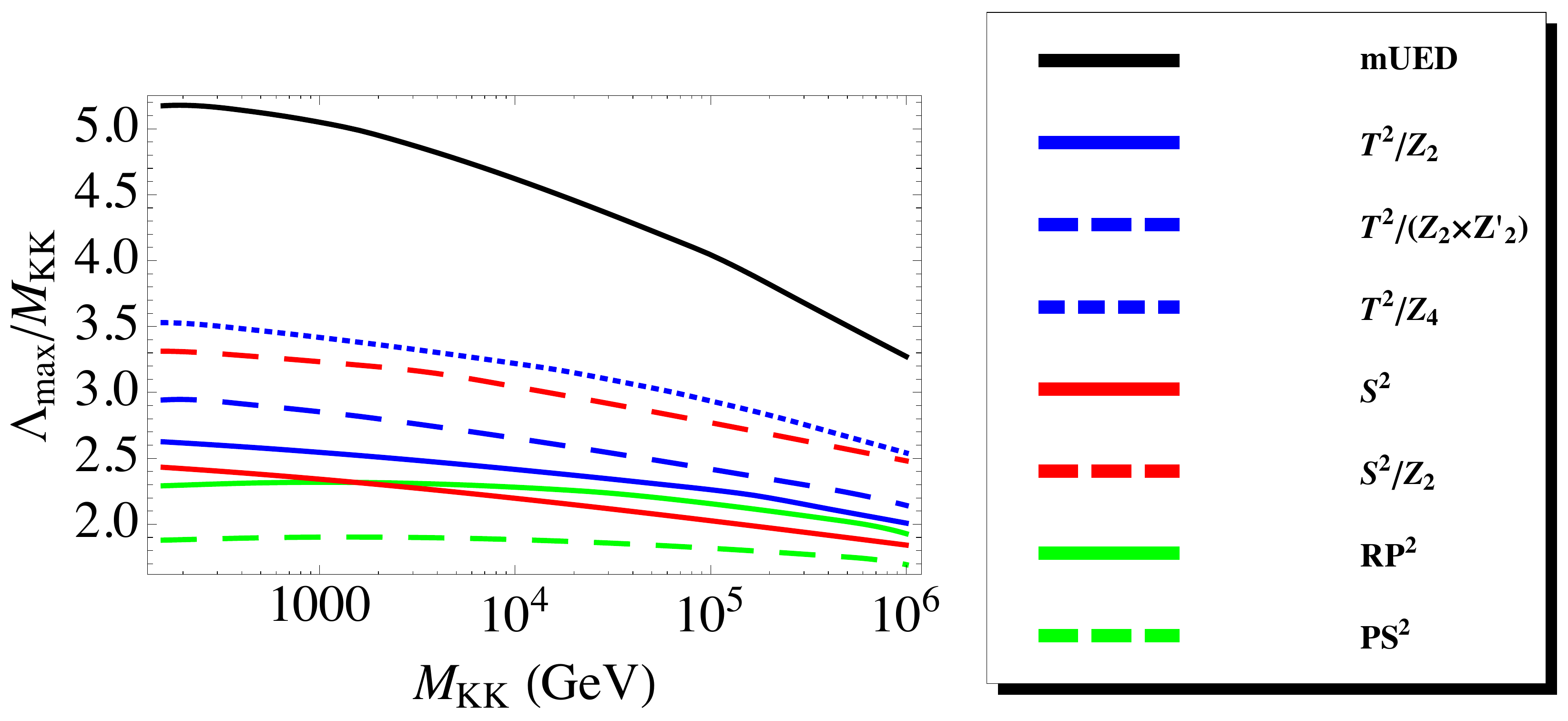}
\caption{{\it Left}\,: {Upper} bounds on the UV cutoff of the UED models as a function of $M_{\text{KK}}$, with the initial conditions in Eq.~(\ref{RGE_initialconditions}).
{\it Right}\,: {Our} color convention for types of the UED models.
The lines in red, blue, and green show the results of $T^2$-based, $S^2$-based, and {nonorientable-manifold-based} UEDs, respectively.}
\label{fig:cutoff_upperbounds}
\end{center}
\end{figure}

\begin{figure}[t]
\begin{center}
   \includegraphics[width=0.6\columnwidth]{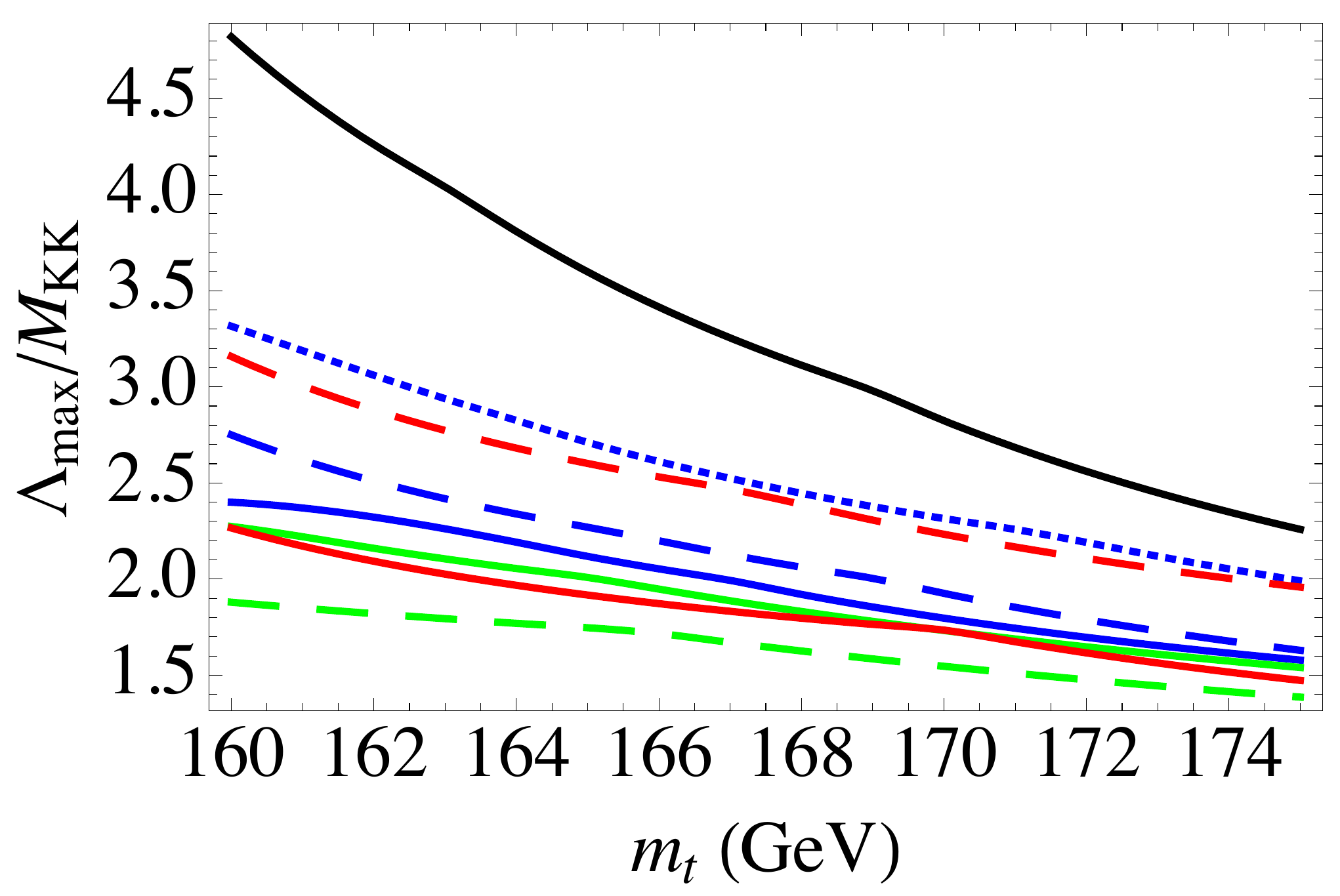}
\caption{Cutoff upper bounds on the 6D UED models and the mUED with the initial condition of the Higgs in Eq.~(\ref{RGE_initialconditions}) and of the top quark in Eq.~(\ref{RGE_initialcondition2}) with {changing} in the region of $[160\,\text{GeV}, 175\,\text{GeV}]$ {with $M_{\text{KK}} = 1\,\text{TeV}$}.
Conventions of line colors and shapes are the same as in Fig.~\ref{fig:cutoff_upperbounds}.
}
\label{fig:cutoff_upperbounds2}
\end{center}
\end{figure}


\begin{table}[t]
\begin{center}
\begin{tabular}{|c||c|c|c|c|c|c|c|c|}
\hline
{Model} &
mUED & 
$T^2/Z_2 $ & 
$T^2/{(Z_2 \times Z'_2)}$ & 
$T^2/Z_4$ & 
$S^2 $ & 
$S^2/Z_2$ & 
$RP^2$ & 
{PS} \\ \hline \hline
$\tilde{\Lambda}_{\text{max}}$ &
5.0 & 2.5 & 2.9 & 3.4 & 2.3 & 3.2 & {2.3} & 1.9 \\ \hline
\end{tabular}
\caption{Upper bounds on cutoff scale $\Lambda_{\text{max}} = \tilde{\Lambda}_{\text{max}} M_{\text{KK}}$ with $M_{\text{KK}} = 1\,\text{TeV}$ and the initial conditions in Eq.~(\ref{RGE_initialconditions}).}
\label{table:maximalcutoff}
\end{center}
\end{table}

Following the discussion in the previous section, we evaluate the constraints on the {highest} {possible} UV cutoff scale $\Lambda$ from vacuum stability of the Higgs potential.
{In our analysis, we literally evaluate the KK summation in Eq.~(\ref{Eq:betafunction}), unlike the previous analysis in Ref.~\cite{Nishiwaki:2011gm} where we obtained the UV cutoff scale of the UED models from the perturbativity of the 4D gauge couplings via the RGEs with its KK summation replaced by an integration.}
In other words, we treat the threshold correction when the reference energy {crosses the} mass of {a} KK particle explicitly in our numerical calculation.
As it was discussed in the previous section, we can ignore the mass coming from the Higgs mechanism with good precision.
Here{,} we adopt the following criterion for determining $\Lambda_{\text{max}}$:
\al{
\lambda(\mu = \Lambda_{\text{max}}) = 0,
\label{cutoff_criterion}
}
where the Higgs potential is destabilized.

We note that {the vacuum stability bound is sensitive to the} differences in the initial condition of the Higgs {self-coupling} $\lambda$ and the top Yukawa coupling $y_t$~\cite{Degrassi:2012ry,Alekhin:2012py}.
In our analysis, we adopt the following values:
\al{
\frac{v^2}{2} \lambda (\mu = m_Z) = 126^2\,\text{GeV}^2, \quad
\frac{v}{\sqrt{2}} y_t (\mu = 173.5\,\text{GeV}) = 160\,\text{GeV},
\label{RGE_initialconditions}
}
where $m_Z$ is the Z-boson mass, the $126\,\text{GeV}$ is the observed Higgs mass at the LHC,
the $173.5\,\text{GeV}$ and the $160\,\text{GeV}$ are the latest values of the pole and the $\overline{\text{MS}}$ masses of the top quark reported by the particle data group~\cite{Beringer:1900zz}, respectively.

The results are summarized in Fig.~\ref{fig:cutoff_upperbounds}{,} and the values with $M_{\text{KK}} = 1\,\text{TeV}$ {are} also {listed} in Table~\ref{table:maximalcutoff}.
$M_{\text{KK}}$ means the first KK mass: $\MKK=1/R$ for the $S^1/Z_2$ (mUED) {and} $T^2$-based compactifications (namely{,} $T^2/Z_2$, $T^2/(Z_2\times Z_2')$, $T^2/Z_4$ and $RP^2$) and ${\MKK}=\sqrt{2}/R$ for the $S^2$-based ones (namely{,} $S^2/Z_2$, PS {and $S^2$}), {where we assume $R_5 = R_6 = R$.}
It is noted that the mUED case has been studied in Refs.~\cite{Bhattacharyya:2006ym,Cornell:2011ge,Blennow:2011tb,Datta:2012db} in many contexts{,} and we find a study in the case of $T^2/Z_4$~\cite{Ohlsson:2012hi}.\footnote{
We can find some related works on the evolutions of {higher-dimensional} neutrino operator in the mUED~{\cite{Bhattacharyya:2002nc,Blennow:2011mp}} and in the $T^2/Z_4$~\cite{Ohlsson:2012hi} {and} of the {Cabibbo--Kobayashi--Maskawa} matrix~\cite{Cornell:2010sz} in the mUED context.
}
We mention that our conclusion {on} the mUED is consistent with that in a previous analysis in Ref.~\cite{Datta:2012db}.
The constraints from vacuum stability, {shown in Table~\ref{table:maximalcutoff},} {are} tighter {than} our previous bounds from perturbativity {of the gauge couplings}: $\Lambda_{\text{max}} \sim 5 M_{\text{KK}}$ in $T^2/Z_2$, $T^2/Z_2 \times Z'_2$, $T^2/Z_4$, $RP^2$ {and} $\Lambda_{\text{max}} \sim 7 M_{\text{KK}}$ in $S^2$, $S^2/Z_2$, {PS}. 
We {note} that{,} in the previous analysis, we ignored differences in types of the compactifications and did not put a bound on the mUED since the KK summations in the single Higgs production and the Higgs decay, which are important in LHC phenomenology and {which} we consider in the next section, are convergent in this case.

Next, we consider the effects when we change the values of top Yukawa coupling in the initial conditions of the RGEs {with $M_{\text{KK}} = 1\,\text{TeV}$}.
We note that{,} within the SM, various values of $\overline{\text{MS}}$ top mass $m_t|_{\overline{\text{MS}}}$ have been reported between {$160$} and $175\,\text{GeV}$~\cite{Beringer:1900zz,Degrassi:2012ry,Alekhin:2012py,Jegerlehner:2012kn}.
Based on this fact, we calculate the bounds on $\lambda$ with varying the initial condition of {the} top Yukawa as
\al{
\frac{v}{\sqrt{2}} y_t (\mu = 173.5\,\text{GeV}) = m_t|_{\overline{\text{MS}}},\quad
\text{for }
160\,\text{GeV} \leq m_t|_{\overline{\text{MS}}} \leq 175\,\text{GeV}.
\label{RGE_initialcondition2}
}

Our result, depicted in Fig.~\ref{fig:cutoff_upperbounds2}, is
sensitive to the value of $m_t|_{\overline{\text{MS}}}$ {and} is consistent with the analyses in {Ref.}~\cite{Blennow:2011tb} (mUED) and in {Ref.}~\cite{Ohlsson:2012hi} ($T^2/Z_4$).
We cannot avoid {the} ambiguity {originating} from {the} top Yukawa coupling.
From Figs.~\ref{fig:cutoff_upperbounds} and \ref{fig:cutoff_upperbounds2}, 
we find that the dependence of $\Lambda$ on $M_{\text{KK}}$ and $m_t|_{\overline{\text{MS}}}$ {is} greater in the mUED than in the 6D UED models.
In the latter, the KK threshold corrections are larger {than those} in the mUED {because of} their denser KK spectra,
and hence the 
{vacuum becomes unstable at a lower energy scale.}

%
%

\section{Higgs signals at Large Hadron Collider
\label{section:LHC}}
{Equipped with} the knowledge for the cutoff scale of UED models in the previous section, 
we estimate the bound on their KK mass scale from the recent results of Higgs search at the LHC.

\subsection{Feature of Higgs signals in UED models}
The structure of the Higgs signal at the LHC {can be} divided into the production and decay. 
The Higgs production is dominated by the gluon fusion process $gg \to H${,} which is induced by the top loop. 
{One of the most important} Higgs decay {channels} that lead to its discovery is the diphoton {one} $H \to \gamma\gamma${,} which is induced by the top and $W$ boson {loops}. 
The Higgs signal is very sensitive to the contribution of the loop corrections at the LHC.
In UED models, a lot of additional KK loops contribute to both $gg \to H (H \to gg)$ and $H \to \gamma\gamma$. 
The KK top loop contribution to the gluon fusion production cross section takes the following form:
\al{
  \label{Eq:GF}   \hat\sigma^{{\text{UED}}}_{gg\to H} &= \frac{\pi^2}{8 m_H} \Gamma ^{{\text{UED}}}_{H\rightarrow gg} \delta (\hat{s} - m_H^2), \\
  \label{Eq:Hgg}   \Gamma ^{{\text{UED}}}_{H\rightarrow gg}  &= K \frac{\alpha^2_s}{8\pi^2}\frac{m_H^3}{v^2_\text{EW}} |J^{{\text{SM}}}_t {(m_H^2)} + J^{{\text{KK}}}_t {(m_H^2)}  |^2, 
}
where {$K\sim1.5$} is {the} {K factor} {accounting} for the {higher-order} QCD corrections {for the case of the LHC}, $\alpha_s = \frac{g_s^2}{4\pi}$ is the fine structure constant for QCD,
$v \simeq$ 246 GeV {is} the electroweak scale, and $J^\text{SM/KK}_t$ {denotes} the SM/KK top quark loop function, defined in {Refs.}~\cite{Nishiwaki:2011gm, Nishiwaki:2011gk}.
The KK top quark and KK {$W$-boson} loop {contributions} to the Higgs decay into diphoton are written as 
\al{
\label{Eq:H2gamma}
\Gamma ^{{\text{UED}}}_{H\rightarrow \gamma \gamma} 
 &= \frac{{\alpha^2} G_F m_H^3}{8\sqrt{2}\pi^3} \left| J^{{\text{SM}}}_W {(m_H^2)} + J^{{\text{KK}}}_W {(m_H^2)} + \frac43 (J^{{\text{SM}}}_t {(m_H^2)} +J^{{\text{KK}}}_t {(m_H^2)}) \right|^2, 
}
where $\alpha = \frac{e^2}{4\pi}$ and $G_F$ are fine structure {constants for the} QED and Fermi constant, {respectively.} 
The SM/KK {$W$-boson loop functions} $J^\text{SM/KK}_W$ are defined in Ref.~\cite{Nishiwaki:2011gk}.
{We {have listed them {in}} Appendix~\ref{Appendix:loopfunction}.}

\begin{figure}[tpbh]
\begin{center}
   \includegraphics[width=40em]{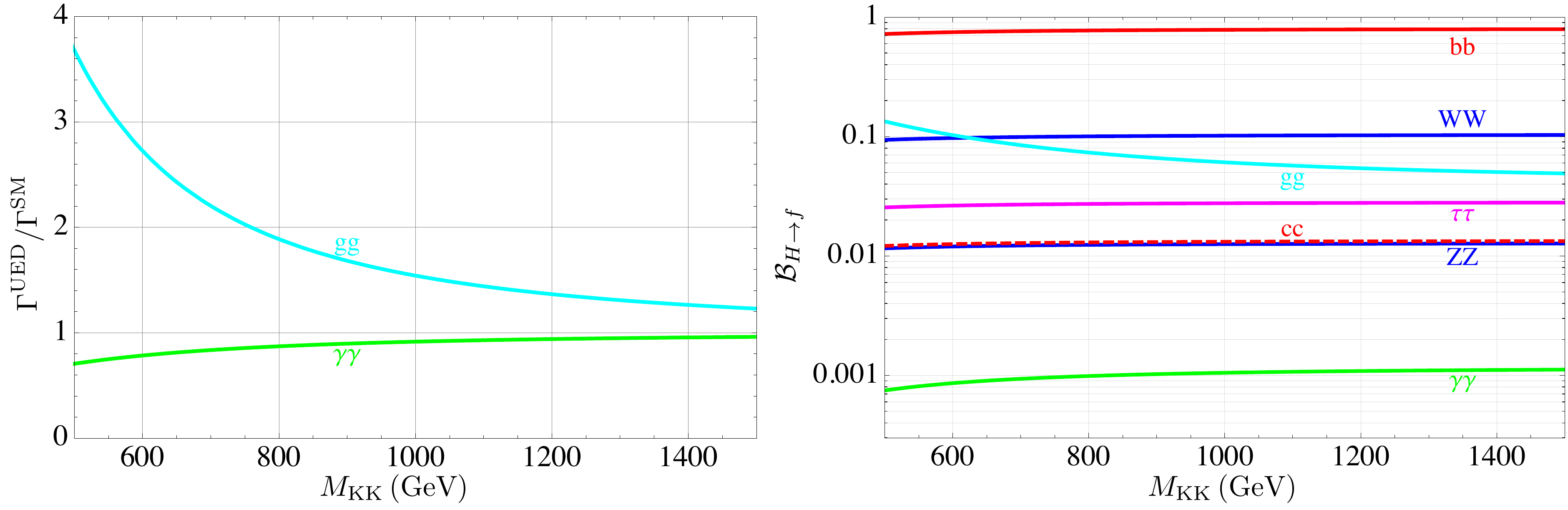}
\caption{
{For the Higgs decay in the $T^2/Z_2$ UED model, we show the UED/SM ratio (left) and the branching ratio (right) as a function of $M_\text{KK}$ for each final state, which is indicated by the caption within Figure; especially we distinguish the almost degenerate $cc$ and $ZZ$ by dashed and solid lines, respectively.}
}
\label{Fig:UEDillustration}
\end{center}
\end{figure}

{Because of} these additional contributions, the {loop-induced} processes $gg \to H$ ($H \to gg$) and $H \to \gamma\gamma$ receive nontrivial effects, 
{which} we {compute and use to estimate} the branching ratios and the Higgs decay {rates} into {the} diphoton and digluon. As {an} illustration, {we show results for {the} $T^2/Z_2$ model} in Fig.~\ref{Fig:UEDillustration}. 
The UED/SM ratio of $H \to gg$ is always enhanced while that of $H \to \gamma\gamma$ is suppressed as already seen in Ref.~\cite{Nishiwaki:2011gk}. 
These behaviors also affect the branching ratios of the Higgs decay as shown in the right panel of Fig.~\ref{Fig:UEDillustration}.
The enhancement in $H \to gg$ is straightforwardly understood as the KK top contributions in the loop diagram. 
The reason {for} the suppression in $H \to \gamma\gamma$ is as follows. 
Since the vectorlike fermions have twice {the} {degrees} of freedom compared with SM fermions, their negative contributions to {the Higgs} decay rate become larger than {the} positive {ones coming} from {the} KK W loop.
{Thus{,} the sum of the KK loops becomes negative, and it overcomes the positive SM contribution.}
As a consequence{,} the decay rate of $H \to \gamma\gamma$ is suppressed compared with {the} SM.

\subsection{Strategy to constrain the KK mass scale}
As shown above, the UED models give different production cross {sections} in the gluon fusion (GF). 
On the other hand, the other productions, which are the vector boson fusion (VBF), the Higgs-strahlung (VH), and the associated production with a $t\bar t$ pair (ttH), are the same as the SM.
{We express the VH production associated with $W$ and $Z$ by WH and ZH, respectively.}
In the recent analysis, the ATLAS and CMS experiments have reported on ratios of these production channels in $H\to \gamma\gamma, ZZ,$ and $WW$ for each category tagging their decays 
\cite{ATLAS:2013gamma,ATLAS:2013Z,ATLAS:2013W,CMS:2013gamma,CMS:2013Z,CMS:2013W}.\footnote{
We use ``$ZZ$'' and ``$WW$'' as the meaning of $ZZ\to 4\ell$ and $WW\to 2\ell2\nu$ for simplicity.}
Such ratios are quite important for obtaining the bound on UED models because of the nontrivial effect of the KK loop corrections on both the production and the decay of the Higgs boson. 
In order to take the different ratios of the production cross section into account, we employ the following quantity~{\cite{Giardino:2012dp,Abe:2012eu}}: 
\begin{equation}
\epsilon_f^{I,X} = \frac{a_f^{I,X} \sigma_X^\text{SM}}{\sum_Y a_f^{I,Y} \sigma_Y^\text{SM}}, 
\end{equation}
where $X$ and $I$ indicate a production channel and a category tagging the decay $H\to f${,} $\sigma_X^\text{SM}$ is the Higgs production cross section of the channel $X$ in the SM{,} 
and $a_f^{I,X}$ is introduced as its acceptance. 
When the set $\{ \epsilon_f^{I,X} \}$ {is} given in the decay $H \to f$, the signal strength {is} written as
\begin{equation}
 \mu^I_{H\to f} =\sum_X \epsilon_f^{I,X}\, \frac{\sigma_X}{\sigma^\text{SM}_X}\, \frac{\mathcal B_{H\to f}}{\mathcal B_{H\to f}^\text{SM}}, 
\end{equation}
where $\sigma_X^\text{(SM)}$ represents the Higgs production cross section of the channel $X${,}
and $\mathcal B_{H\to f}^\text{(SM)} =\Gamma_{H\to f}^\text{(SM)} / \Gamma_{H\to \text{all}}^\text{(SM)}$ is the branching ratio of the Higgs decay $H \to f$ (in the SM). 
In the UED model, $\sigma_\text{GF} = \hat\sigma_{gg\to H}^\text{UED}$,\, $\Gamma_{H\to \gamma\gamma (gg)} =\Gamma_{H\to \gamma\gamma (gg)}^\text{UED}$ as in Eqs.~(\ref{Eq:GF})--(\ref{Eq:H2gamma}){,}
and the others are assumed to be the same as the SM in our analysis. 

\begin{table}[tpbh]
\begin{center}
\begin{tabular}{c|c|ccccc}
\hline\hline
              $I$                         & $\mu^I_{H\to \gamma\gamma}$ &             \multicolumn{5}{c}{$\epsilon_{\gamma\gamma}^{I,X}$ (\%) }       \\ \hline
{Event} category                  &   {Signal} strength                         &   GF  &    VBF  &   VH(WH)   &   VH(ZH)  &  ttH \\ \hline
{Unconventional} central low  $P_T$&   $0.9\pm0.7 $   &   93.7   &   4.0     &  1.4     & 0.8     &   0.2   \\ \hline
{Unconventional} central high $P_T$&   $1.0^{+1.1}_{-0.9}  $  &    79.3  &   12.6    &  4.1     &  2.5   &   1.4    \\ \hline
{Unconventional} rest low $P_T$    &   $2.6^{+0.9}_{-1.0}  $  &    93.2  &   4.0      &  1.6     &  1.0   &   0.1    \\ \hline
{Unconventional} rest high $P_T$   &   $2.7^{+1.3}_{-1.2}  $  &    78.1  &   13.3     &  4.7     & 2.8   &   1.1    \\ \hline
{Conventional} central low $P_T$   &   $1.4^{+1.0}_{-0.9}  $  &    93.6  &   4.0      &  1.3     & 0.9   &   0.2    \\ \hline
{Conventional} central high $P_T$  &   $2.0^{+1.5}_{-1.3}  $  &    78.9  &   12.6     &  4.3     & 2.7   &   1.5    \\ \hline
{Conventional} rest low  $P_T$     &   $2.2^{+1.2}_{-1.0}  $  &    93.2  &   4.1     &  1.6      & 1.0   &   0.1    \\ \hline
{Conventional} rest high $P_T$     &   $1.3\pm1.3  $  &    77.7  &   13.0     &  5.2      & 3.0   &   1.1    \\ \hline
{Conventional} transition          &   $2.8^{+1.7}_{-1.6}  $  &   90.7   &   5.5     &   2.2     & 1.3    &  0.2    \\ \hline
Loose high mass 2 jet     &   $2.8^{+1.7}_{-1.4}  $  &    45.0  &   54.1     &  0.5     & 0.3   &   0.1    \\ \hline
Tight high mass 2 jet     &   $1.6^{+0.8}_{-0.6}  $  &    23.8  &   76.0     &  0.1     & 0.1   &   0.0    \\ \hline
Low mass 2 jet            &   $0.3^{+1.7}_{-1.5}  $  &    48.1  &   3.0     &  29.7     & 17.2   &   1.9    \\ \hline
$E^{miss}_T$ significance   &   $3.0^{+2.7}_{-1.9}  $  &    4.1   &   0.5     &  35.7     & 47.6   &   12.1   \\ \hline
One lepton                &   $2.7^{+2.0}_{-1.7}  $  &    2.2  &   0.6      &  63.2     & 15.4   &   18.6    \\ \hline\hline
\end{tabular} 
\caption{The ATLAS result of $H\to\gamma\gamma$ analysis. 
The ATLAS experiment defines these event categories and uses these ratios of the production channels as in Ref.~\cite{ATLAS:2013gamma}. }
\label{Tab:ATLASgamma}
\end{center}
\end{table}
\begin{table}[tpbh]
\begin{center}
\begin{tabular}{c|c|cccc}
\hline\hline
              $I$                         & $\mu^I_{H\to \gamma\gamma}$ &             \multicolumn{4}{c}{$\epsilon_{\gamma\gamma}^{I,X}$ (\%) }       \\ \hline
{Event} category         &   {Signal} strength          &   GF  &    VBF  &   VH   & ttH    \\ \hline
Missing {$E_T$}             &   $1.9^{+2.6}_{-2.3} $   &   22.0   &   2.6     &  63.7     &  11.7     \\ \hline
{Electron} tag             &   $-0.7^{+2.8}_{-2.0}  $  &   1.1   &   0.4     &  78.7     &  20.8     \\ \hline
muon tag                 &   $0.4^{+1.8}_{-1.4}  $  &    0     &   0.2      & 79.0    &   19.8    \\ \hline
2-jet loose              &   $0.8^{+1.1}_{-1.0}  $  &    47.0  &   50.9     &  1.7     &  0.5     \\ \hline
2-jet tight              &   $0.3^{+0.7}_{-0.6}  $  &   20.7  &  78.9      &  0.3     &   0.1       \\ \hline
Untag-3                  &   $-0.3^{+0.8}_{-0.9}  $  &   92.5  &   3.9      &  3.3     &   0.3    \\ \hline
Untag-2                  &   $0.3\pm0.5  $  &    91.6  &   4.5      &  3.6     &   0.4    \\ \hline
Untag-1                  &   $0.0\pm0.7  $  &    83.5  &  8.4      &  7.1      &   1.0    \\ \hline
Untag-0                  &   $2.2^{+0.9}_{-0.8}  $  &    72.9  &   11.6     &  12.9     &   2.6    \\ \hline
2-jet {(7\,TeV)}               &   $4.2^{+2.3}_{-1.8}  $  &    26.8  &   72.5    &  0.6     &   {0}    \\ \hline
Untag-3 {(7\,TeV)}            &   $1.5^{+1.7}_{-1.8}  $  &    91.3  &   4.4      &  4.1      &  0.2    \\ \hline
Untag-2 {(7\,TeV)}            &   $0.0^{+1.3}_{-1.2}  $  &    91.3  &   4.4     &  3.9     &   0.3   \\ \hline
Untag-1 {(7\,TeV)}            &   $0.2^{+1.0}_{-1.0}  $  &    87.6    &  6.2      & 5.6     &  0.5    \\ \hline
Untag-0 {(7\,TeV)}            &   $3.8^{+2.0}_{-1.7}  $  &    61.4    & 16.8      & 18.7     & 3.1    \\ \hline\hline
\end{tabular} 
\caption{The CMS result of $H\to\gamma\gamma$ analysis. 
The CMS experiment defines these event categories and uses these ratios of the production channels as in Ref.~\cite{CMS:2013gamma}. }
\label{Tab:CMSgamma}
\end{center}
\end{table}
\begin{table}[tpbh]
\begin{center}
\begin{tabular}{c|c|cc}
\hline\hline
          $I$             &  $\mu^I_{H\to ZZ}$            & \multicolumn{2}{c}{$\epsilon_{ZZ}^{I,X}$ (\%) } \\ \hline
{Event} category    &   {Signal} strength                 & GF                    & VBF          \\ \hline
Untagged            &   $0.85^{+0.32}_{-0.26}  $  & 95                     & 5               \\ \hline
{2-jet} tag               &   $1.22^{+0.84}_{-0.57}  $  & 80                     & 20             \\ \hline\hline
          $I$             &  $\mu^I_{H\to WW}$             & \multicolumn{2}{c}{$\epsilon_{WW}^{I,X}$ (\%) } \\ \hline
SF 1 jet {(7\,TeV)}  &   $0.9^{+2.1}_{-2.2}  $        & 100                    & {0}                  \\ \hline
SF 0 jet {(7\,TeV)}  &   $0.1\pm1.0  $                   & 100                    & {0}                  \\ \hline
DF 1 jet {(7\,TeV)}   &   $1.7\pm1.0  $                   & 100                    & {0}                \\ \hline
DF 0 jet {(7\,TeV)}   &   $0.6\pm0.5  $                 & 100                    & {0}                  \\ \hline
SF 1 jet {(8\,TeV)}   &   $1.5\pm0.9  $                   & 100                    & {0}                  \\ \hline
SF 0 jet {(8\,TeV)}   &   $1.1\pm0.7  $                   & 100                    & {0}                  \\ \hline
DF 1 jet {(8\,TeV)}   &   $0.3\pm0.4  $                   & 100                    & {0}                  \\ \hline
DF 0 jet {(8\,TeV)}   &   $0.7\pm0.3  $                   & 100                    & {0}                  \\ \hline\hline
\end{tabular} 
\caption{The CMS result of $H\to ZZ/WW$ analysis. 
The CMS experiment defines these event categories and uses these ratios of the production channels as in Ref.~\cite{CMS:2013Z,CMS:2013W}.
{SF and DF denote ``same flavor'' and ``different flavor'', respectively.}}
\label{Tab:CMSzwboson}
\end{center}
\end{table}
%
For the analysis in $H\to \gamma\gamma$, the ATLAS and CMS experiments have shown their results of $\mu^I_{H\to \gamma\gamma}$ 
and the set $\{ \epsilon_{\gamma\gamma}^{I,X} \}$ they used in their analyses~\cite{ATLAS:2013gamma, CMS:2013gamma}.    
We summarize these values in {Tables}~\ref{Tab:ATLASgamma} and \ref{Tab:CMSgamma}.
For the analysis in $H\to ZZ/WW$, the CMS result is summarized in Table~\ref{Tab:CMSzwboson}. 
The result of $H\to WW$ in the CMS experiment is given by assuming that all Higgs signals are produced by the GF process~\cite{CMS:2013W}.  
The ATLAS experiment only gives the signal strength for the specific production channels~\cite{ATLAS:2013Z, ATLAS:2013W},  {which} is written as 
\begin{equation}
\mu^X_{H\to ZZ/WW} = \frac{\sigma_X}{\sigma^\text{SM}_X}\, \frac{\mathcal B_{H\to ZZ/WW}}{\mathcal B_{H\to ZZ/WW}^\text{SM}} . 
\end{equation}
The results are given as $\mu^\text{GF+ttH}_{H\to ZZ}= 1.8^{+0.8}_{-0.5}$, $\mu^\text{VBF+VH}_{H\to ZZ}= 1.2^{+3.8}_{-1.4}$, $\mu^\text{GF}_{H\to WW}= 0.82\pm0.36${,} 
and $\mu^\text{VBF}_{H\to WW}= 1.66\pm0.79$. 
In this article, we assume $\mu^\text{GF+ttH}_{H\to ZZ} \simeq \mu^\text{GF}_{H\to ZZ}$ for simplicity.

We evaluate a bound on the KK scale in {each UED model} by performing {a} $\chi^2$ analysis of the results as shown above. 
The $\chi^2$ function is represented as 
\begin{equation}
 \chi^2 =\sum_f \sum_I \left( \frac{ \mu^I_{H\to f} - \hat\mu^I_f }{\hat\sigma^I_f} \right)^2,
\end{equation}
where we assume the experimental results to be Gaussian distribution $\hat\mu^I_f \pm \hat\sigma^I_f$.\footnote{
{Note that{,} since we neglect the correlation among the categories, which is not made public, this analysis should rather be taken as an illustration.}
}
The number of the observables we use in our analysis is 42 in total{,} and the degree of freedom is also the same number in terms of testing a justification of a model.



\subsection{Bound on KK scale from the current data }
\begin{figure}[tpbh]
\begin{center}
   \includegraphics[viewport=0 0 725 458, width=32em]{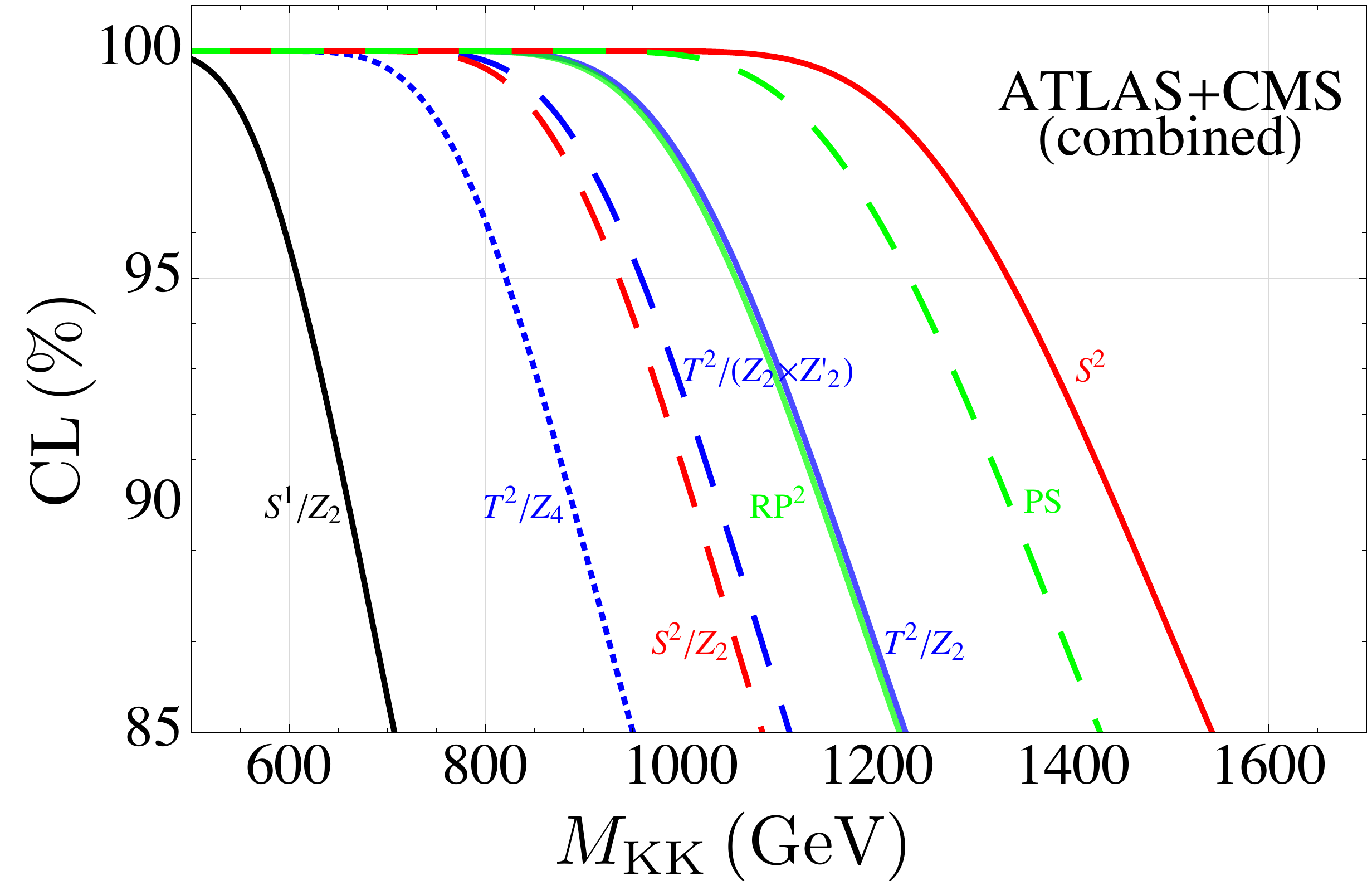}
\caption{
{
Exclusion {C.L.s} of all the UED models as functions of the KK scale $M_\text{KK}$ by use of all the ATLAS and CMS results of $H \rightarrow \gamma \gamma,WW,ZZ$. 
Colors denote the same as in {Fig.}~\ref{fig:cutoff_upperbounds}. 
}
}
\label{Fig:BoundHiggs}
\end{center}
\end{figure} 
Here we show bounds on several UED models from the Higgs searches at the LHC. 
{For our analyses, we have taken the highest possible UV cutoff scale $\Lambda_\text{max}$ shown in Table~\ref{table:maximalcutoff}. The Higgs mass is chosen to be 126\,GeV.}
In Fig.~\ref{Fig:BoundHiggs}, we show the exclusion {C.L.} of each UED model as a function of the KK scale $M_\text{KK}$ 
by use of all the ATLAS and CMS results of $H\to \gamma\gamma, WW, ZZ$. 
The black line {indicates} the result in the {five-dimensional} mUED model. 
The blue {solid}, dashed, and dotted {lines} denote {those} in the {$T^2$-based ones, namely{,} the} $T^2/Z_2$, {$T^2/(Z_2 \times Z'_2)$} and $T^2/Z_4$, respectively. 
The red {solid} and dashed line{s} represent {those in the $S^2$-based ones, namely} $S^2$ and $S^2/Z_2$, respectively. 
The green {solid} and dashed line{s} show {those} in the {nonorientable} ones, namely{,} $RP^2$ and PS, respectively.

As {can be} seen in this graph, we find that the region $M_\text{KK} \lesssim {600}$ GeV is excluded with{in} $95\%$ {C.L.} in the mUED model. 
For the {six-dimensional} models {in the} $T^2$-base{d space}, 
we find the {excluded} regions $M_\text{KK} \lesssim 1100$, ${1000}${,} and ${800}$ GeV with{in} $95\%$ {C.L.} {for} $T^2/Z_2$, $T^2/(Z_2 \times Z'_2)${,} and $T^2/Z_4$, respectively. 
For the $S^2$-based models, 
we can see that the regions $M_\text{KK} \lesssim {1300}$ and ${900}$ GeV are excluded with{in} $95\%$ {C.L.} in the $S^2$ and $S^2/Z_2$, respectively. 
For the {nonoriented} models, 
the regions $M_\text{KK} \lesssim {1100}$ and ${1200}$ GeV are excluded with{in} $95\%$ {C.L.} in the $RP^2$ and PS,  respectively. 
As seen above, the excluded region is different from one model to another in the case for UED. 
This is because the difference of the KK spectrum has a large impact on the Higgs decays via loop processes.

\begin{figure}[tpbh]
\begin{center}
   \includegraphics[viewport=0 0 725 458, width=18em]{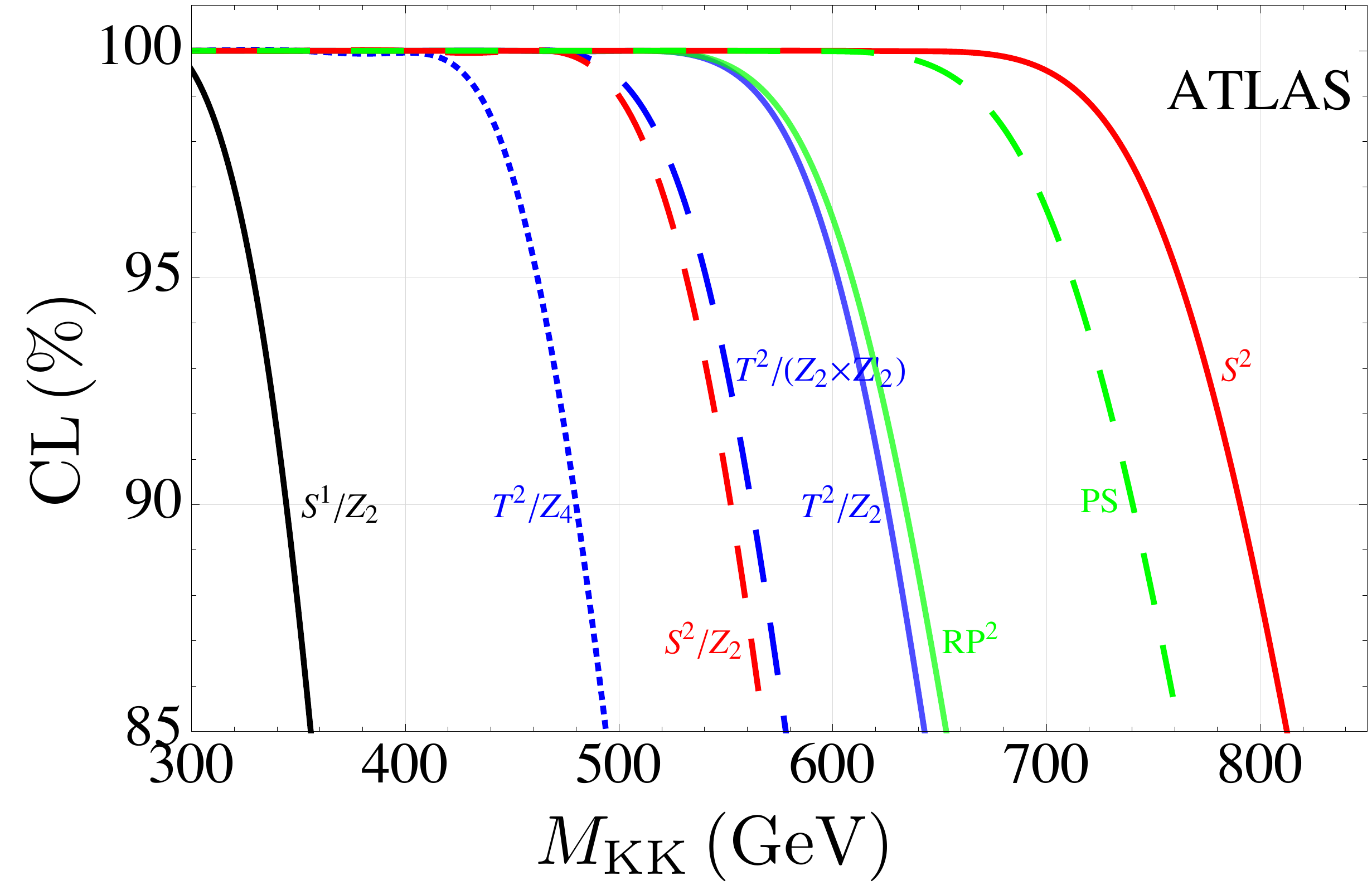}
   \includegraphics[viewport=0 0 725 458, width=18em]{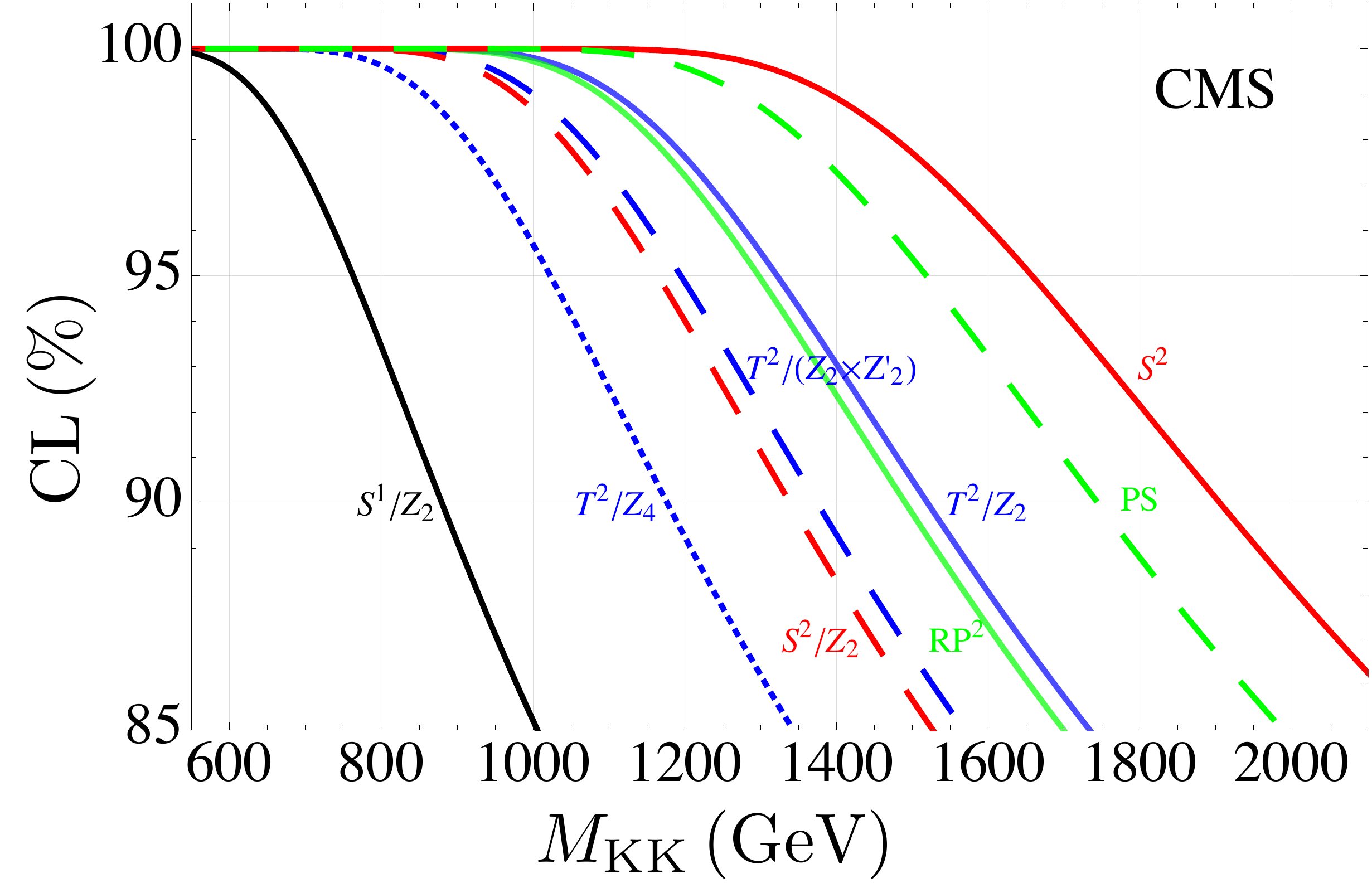}
\caption{{The exclusion {C.L.s} of all UED models as functions of the KK scale $M_\text{KK}$ obtained from the ATLAS (left) and CMS (right) results of $H\to \gamma\gamma, WW, ZZ$. 
Colors denote the same as in {Fig.}~\ref{fig:cutoff_upperbounds}. 
}
}
\label{Fig:BoundHiggsCMSandATLAS}
\end{center}
\end{figure} 
We compare the bounds obtained from the ATLAS experiment with {those from} the CMS in Fig.~\ref{Fig:BoundHiggsCMSandATLAS}.  
We find that the CMS result gives {a} more stringent bound on the KK scale compared with the ATLAS {one.} 
In other words, for now{,} the UED models are likely to explain the recent ATLAS result, while they are disfavored by the recent CMS result. 
The results of the exclusion {C.L.} for the wide range of the KK scale are summarized in {Fig.~\ref{Fig:UEDboundWide} in Appendix~\ref{Appendix:bounds}.}

{Throughout this analysis, we ignore the effects from the {higher-dimensional} operators around~$\Lambda$.} {See Ref.~\cite{Nishiwaki:2011vi} for such an effect.}


\section{Indirect constraint from $S$ and $T$ parameters
\label{section:ST}}

Physics beyond the SM is also restricted through the precise measurement of some electroweak {variables}.
The $S$ and $T$ parameters proposed by Peskin and Takeuchi~\cite{Peskin:1990zt,Peskin:1991sw} have been used for estimating whether a model is valid or not.
The {variables}  are defined by use of the two-point functions of the SM gauge bosons,
\al{
\Pi^{\mu\nu}_{ab} (k) =  i \Pi^{\text{T}}_{ab} (k^2) \left( g^{\mu\nu} - \frac{k^{\mu}k^{\nu}}{k^2} \right) + i \Pi^{\text{L}}_{ab} (k^2) \frac{k^{\mu}k^{\nu}}{k^2} \label{gauge_twopointfunction},
}
where $k$ is the external momentum, T (L) means that it is {the} transverse (longitudinal) part, and the indices $a,b$ show types of the SM gauge bosons.
The {variables}  are constructed by the transverse components{,} and the concrete forms are written down with adapting the notation on the electroweak sector in~{Ref.}~\cite{Denner:1991kt} as
\al{
\frac{\alpha S}{4 s_W^2 c_W^2} &=
		{\Pi^{\text{T}}_{ZZ}}'(0) + \frac{c_W^2 - s_W^2}{c_W s_W} {\Pi_{Z\gamma}^{\text{T}}}'(0)
		- {\Pi_{\gamma \gamma}^{\text{T}}}'(0), \\
\alpha T &= \frac{\Pi_{WW}^{\text{T}}(0)}{m_W^2} - \frac{\Pi_{ZZ}^{\text{T}}(0)}{m_Z^2}
		+ {2 c_W s_W} \frac{\Pi_{Z\gamma}^{\text{T}}(0)}{m_W^2},
} 
where ${\Pi^{\text{T}}_{ab}}'$ is defined as $\frac{d}{dk^2} \Pi^{\text{T}}_{ab}(k^2)$.

The $S$ and $T$ are also described by combinations of some electroweak {variables}  and their values are calculated in global analysis with experimental results.
One of the latest numbers {is} found in {Ref.}~\cite{Baak:2012kk},
\al{
S|_{U=0} = 0.05 \pm 0.09,\quad
T|_{U=0} = 0.08 \pm 0.07,\quad
\rho_{ST} = +0.91,
\label{ST_experimental}
}
with {$126\,\text{GeV}$ reference Higgs mass and} assuming the $U$ parameter {is} zero and $\rho_{ST}$ is the correlation coefficient.
In an operator-analysis point of view, the $U$ parameter is represented as a coefficient of a much {higher-dimensional} operator with the Higgs doublet compared with $S$ and $T$ in the UED models, and hence we ignore the effect in our analysis.

\subsection{Forms in 6D UED models and mUED}

In this section, we formulate the contributions to the $S$ and $T$ parameters in the 6D UED models and in the mUED model.
It is well known that the $S$ and $T$ parameters are logarithmically divergent in six dimensions~\cite{Appelquist:2002wb}.
To have a rough idea of what happens, we employ the following prescription.
First{,} we compute the contributions from each KK mode within {four-dimensional} field theory employing the dimensional regularization. They are manifestly finite. Then we sum such contributions up to a mode {in which} the KK mass exceeds the UV cutoff $\Lambda$. To estimate the possible effects from the UV theory above $\Lambda$, we also put the {higher-dimensional} operators in six dimensions.

The general shape of the $S$ and $T$ parameters are
\al{
S = \sum_{s \atop \text{with }M_s < \Lambda} \left( S^{(\text{KK})}_{s,\text{boson}} + S^{(\text{KK})}_{s,\text{fermion}}  \right) + S_{\text{Higgs calibration}} + S_{\text{threshold}}, \label{Sform} \\
T = \sum_{s \atop \text{with }M_s < \Lambda} \left( T^{(\text{KK})}_{s,\text{boson}} + T^{(\text{KK})}_{s,\text{fermion}}  \right) + T_{\text{Higgs calibration}} + T_{\text{threshold}}, \label{Tform}
}
where the first terms {in parentheses are} the contributions of KK particles{,} and the last two terms represent, respectively, the effects from Higgs mass calibration and the threshold correction via possible {higher-dimensional} operators around the UV cutoff scale $\Lambda$ in six dimensions.
These effects were considered in {Refs.}~\cite{Peskin:1991sw,Appelquist:2002wb}{,}
\al{
S_{\text{Higgs calibration}} &= \frac{1}{12\pi} \ln\left( \frac{m_H^2}{m_{H,\text{ref}}^2} \right), &
T_{\text{Higgs calibration}} &= - \frac{3}{12\pi c_W^2} \ln\left( \frac{m_H^2}{m_{H,\text{ref}}^2} \right), \\
S_{\text{threshold}} &= c_S \frac{2 \pi v^2}{\Lambda^2}, &
T_{\text{threshold}} &= c_T \frac{m_H^2}{4 \alpha \Lambda^2},
\label{ST_threshold}
}
where $m_{H,\text{ref}}$ is the assumed SM Higgs mass in global analysis{,} and $c_S$ and $c_T$ are undetermined dimensionless coefficients with $\mathcal{O}(1)$ magnitude.

{Several comments are in order.}
One is that the summations over KK states are truncated at the scale $\Lambda$.
The other is that the value of $\Lambda$ is estimated through the vacuum stability condition of the Higgs boson.
{We choose the highest possible $\Lambda$ allowed by it.}
As we discussed in {Sec.}~\ref{section:2}, in the configuration of $m_H = 126\,\text{GeV}$, the value of the {maximum UV} cutoff scale tends to be low{,} and the threshold corrections possibly become important. We {will} include these effects {below.} 
Finally, we comment on the contributions of KK particles.
We find that the effect from the state-$s$ fermion loops {takes} the following general shapes in every 6D UED {model}, which is the same as in the mUED and was already calculated in {Ref.}~\cite{Appelquist:2002wb}. We show them in our notation:
\al{
S_{\text{fermion},s}^{(\text{KK})} &\simeq \frac{1}{4\pi} \frac{2}{3} x_{t,s}, \quad
T_{\text{fermion},s}^{(\text{KK})} \simeq \frac{1}{\alpha} \left( \frac{m_t^2}{4\pi^2 v^2} \right) x_{t,s},
\label{fermionST}
}
where $x_{t,s}$ is defined with the KK mass of the state $``s"$ $M_s$ as
\al{
x_{t,s} = \frac{m_t^2}{M_s^2}
\label{xoftop}
}
and we ignore their $\mathcal{O}(x_{t,s}^2)$ corrections.
In the $RP^2$ model, we should pay attention to the fact that the summation range differs between bosonic and fermionic sector{s}.

The bosonic part is highly {model dependent}. 
{In this paper,} we {have} newly calculated the contributions to $S$ and $T$ in every 6D model.
The complete forms of the gauge-boson two-point functions are summarized in Appendix~\ref{Appendix:twopoint}.

In the cases of $T^2/Z_2$, {$T^2/(Z_2 \times Z'_2)$}, $T^2/Z_4$, $S^2$, and $S^2/Z_2$, the forms are 
\al{
S_{\text{boson},s}^{(\text{KK})} &\simeq \frac{1}{\pi} \Bigg\{ -\frac{5}{36} x_{W,s} + \frac{1}{24} x_{H,s}
		+ \left( \frac{1}{24} - \frac{1}{6 c_W^2} \right) x_{Z,s} \Bigg\}, \\
T_{\text{boson},s}^{(\text{KK})} &\simeq \frac{1}{4\pi} \frac{1}{s_W^2} \Bigg\{
		\left( \frac{15}{4} - \frac{193}{72} \frac{1}{c_W^2} + \frac{1}{2 c_W^4} \right) x_{W,s} +
		\left( - \frac{85}{72} - \frac{7}{18} \frac{1}{c_W^2} \right) x_{Z,s} +
		\left( \frac{13}{36} + \frac{5}{36} \frac{1}{c_W^2} - \frac{1}{2 c_W^4} \right) x_{H,s} \Bigg\}, 
}
where we define similar variables as in Eq.~(\ref{xoftop}): $x_{i,s} = \frac{m_i^2}{M_s^2}$ with
$m_W^2$, $m_Z^2$, and $m_H^2$. 
{Note} that the lighter the KK particles are the greater they contribute to $S$ and $T$.
{In these models,} the result is affected only by the differences in the patterns of the {surviving} KK {modes}.

In the cases of the models based on the {nonorientable} manifolds $RP^2$ and {PS},
bosonic contributions are classified into {three and two {categories}, respectively}.
The details of the following classifications have already been discussed in {Sec.}~\ref{section:2} and thus we do not explain it here.
The results in the {PS} model are shown{:}
\al{
S_{\text{boson},s:\text{odd}}^{(\text{KK})} &\simeq 0, \quad
T_{\text{boson},s:\text{odd}}^{(\text{KK})} \simeq \frac{1}{4\pi} \frac{1}{s_W^2} \frac{5}{18} \Bigg\{
		\left( 1 - \frac{1}{c_W^2} \right) x_{W,s} +
		\left( \frac{1}{c_W^2} -1 \right) x_{Z,s} \Bigg\},
	    \\
S_{\text{boson},s:\text{even}}^{(\text{KK})} &\simeq \frac{1}{\pi} \Bigg\{ -\frac{5}{36} x_{W,s} + \frac{1}{24} x_{H,s}
		+ \left( \frac{1}{24} - \frac{1}{6 c_W^2} \right) x_{Z,s} \Bigg\},
		\label{Sboson_even} \\
T_{\text{boson},s:\text{even}}^{(\text{KK})} &\simeq \frac{1}{4\pi} \frac{1}{s_W^2} \Bigg\{
		\left( \frac{125}{36} - \frac{173}{72} \frac{1}{c_W^2} + \frac{1}{2 c_W^4} \right) x_{W,s} +
		\left( - \frac{65}{72} - \frac{2}{3} \frac{1}{c_W^2} \right) x_{Z,s} +
		\left( \frac{13}{36} + \frac{5}{36} \frac{1}{c_W^2} - \frac{1}{2 c_W^4} \right) x_{H,s} \Bigg\}.
		\label{Tboson_even}
}
The shapes in the $RP^2$ model are closely related the previous ones in the {PS} as follows:
\al{
\{S,\,T\}_{\text{boson},s:\text{region I}}^{(\text{KK})} &= 
\{S,\,T\}_{\text{boson},s:\text{even}}^{(\text{KK})}, \\
\{S,\,T\}_{\text{boson},s:\text{region II}}^{(\text{KK})} &= 
\{S,\,T\}_{\text{boson},s:\text{odd}}^{(\text{KK})}, \\
\{S,\,T\}_{\text{boson},s:\text{region III}}^{(\text{KK})} &= 
\{S,\,T\}_{\text{boson},s}^{(\text{KK})},
}
where we note that we should use the form of {the} KK mass on $S^2$ instead of on $T^2$.

The mixing among KK states in the gauge sector is schematically of the form
\al{
\begin{bmatrix}
m_{W,Z}^2 + M_\text{KK}^2 & M_\text{KK}^2 & m_{W,Z} M_\text{KK} \\
M_\text{KK}^2 & m_{W,Z}^2 + M_\text{KK}^2 & m_{W,Z} M_\text{KK} \\
m_{W,Z} M_\text{KK} & m_{W,Z} M_\text{KK} & m_{W,Z}^2 + M_\text{KK}^2
\end{bmatrix}.
}In the calculation of $S$ and $T$ parameters, we adopt the following approximation about the mass mixings of 6D W and Z boson-related sectors:
\begin{itemize}
\item We ignore off-diagonal terms with the magnitude $\mathcal{O}(m_{W,Z} M_{\text{KK}})$, which are small compared with the other terms with the magnitude $\mathcal{O}(M_{\text{KK}}^2)$.
\item In the diagonal terms, {for which the} forms are approximately as {${m_{W,Z}^2 + M_{\text{KK}}^2}$}, we do not ignore the small part coming from $m_{W,Z}^2$ since this part can contribute to the $T$ parameter.
\end{itemize}
{Because of} this approximation, the small mixings being proportional to $m_{W,Z}^2$ are ignored.
As a result, some divergent terms that are proportional to $m_{W,Z}^2$ {remain} in the $T$ parameter{,} and we simply discard them.
Note that the contribution to $T$ from each KK mode must be manifestly finite since it is computed in {four-dimensional} field theory with the dimensional regularization. The divergence ($\propto m_{W,Z}^2$) that we encounter here is an artifact coming from the ignorance of the small off-diagonal part in the KK mixing of gauge sector.
Indeed, we find that there appears no divergence proportional to $m_{H}^2$ or $m_{t}^2$, as we treat the mixing of the Higgs and top KK sectors exactly.
{Although} there might be a further possible finite correction due to this procedure, the KK gauge contribution is generally {subleading} compared to the KK top loops, {the mixing of which} we treat exactly.
In the $S$ parameter, we do not see any divergence even after the above approximation.
These features are consistent with the general property of $S$ and $T$.

After we considered radiative corrections, the Weinberg angles of the KK $W$ and $Z$ bosons get to be very small~\cite{Cheng:2002iz}.
We {assume} in this effect {that} the KK Weinberg angles {are} zero and {that} we {can} simply ignore the mass corrections.
Each KK-state {contribution} should be suppressed by {its} KK mass, and hence
this effect {should} not affect the leading order of $S$ and $T$ 
since their contributions are proportional to KK masses (when we {ignore} the electroweak masses in loop calculation)~\cite{Cheng:2002iz}.

Finally we comment on the mUED {model}, {which has been studied extensively}~\cite{Appelquist:2000nn,Appelquist:2002wb,Flacke:2005hb,Gogoladze:2006br,Baak:2011ze}.
In the $\chi^2$ {analysis} of Refs.~\cite{Appelquist:2002wb,Baak:2011ze}, the authors simply ignored the terms being proportional to $m_{W,Z}^2$, possibly {because} their effects are not significant compared with those {that} are proportional to $m_H^2$ or $m_t^2$.

Boson contributions to $S$ and $T$ in the {mUED} are approximately described with the even part of the {PS} (or region I of the $RP^2$) as we already showed in Eqs.~(\ref{Sboson_even}) and (\ref{Tboson_even}) since the particle content of each KK state is the same.
Fermion contributions are the same as in Eq.~(\ref{fermionST}).
Here{,} we adopt the form of the KK mass $M_{(n)}$ with a KK number $n$
\al{
M_{(n)}^2 = {n^2}{M_{\text{KK}}^2},\quad
(\text{for } n=1,2,3,\cdots).
}

\subsection{Numerical results without threshold correction}

\begin{figure}[t]
\begin{center}
   \includegraphics[width=32em]{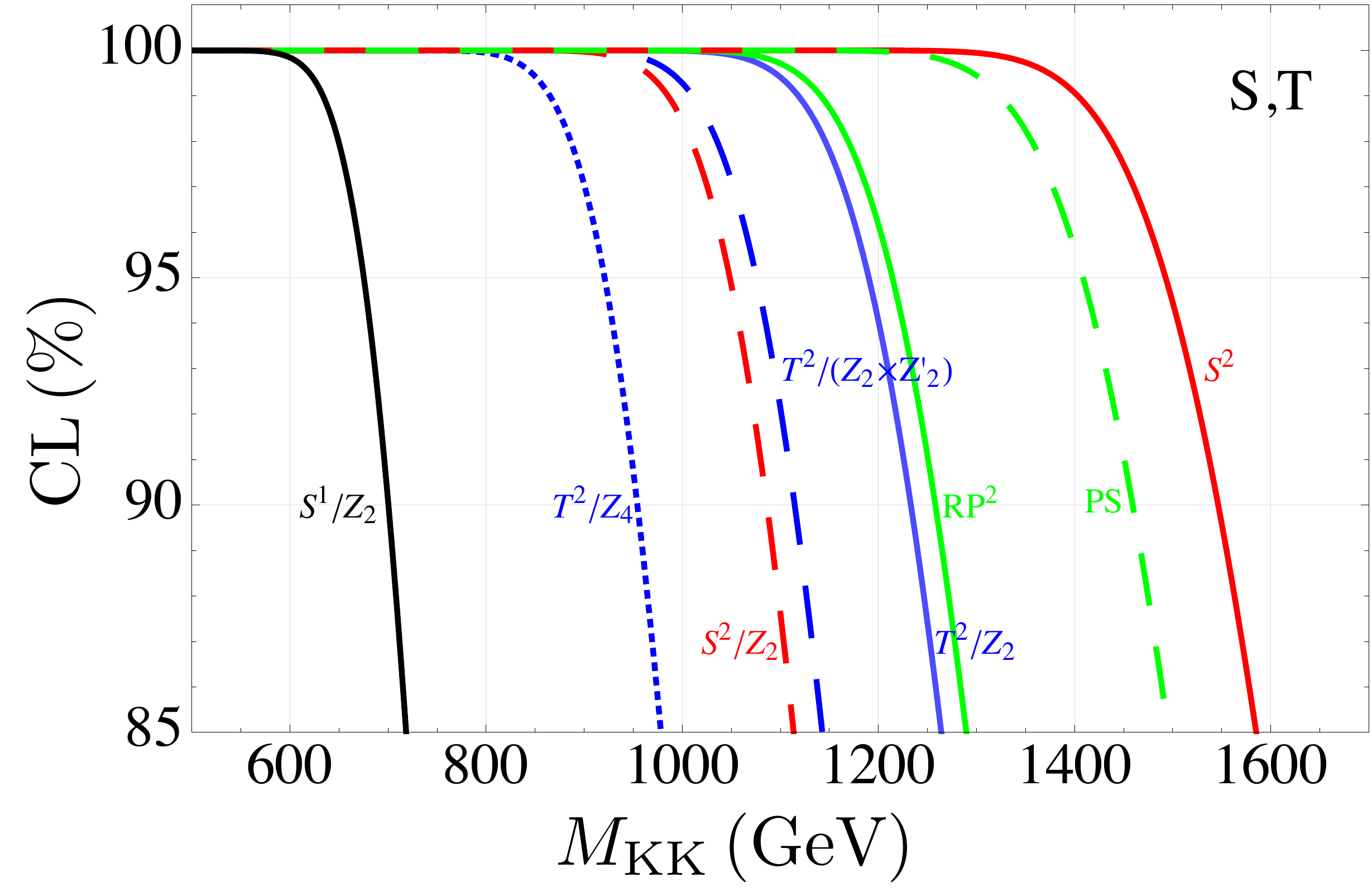}
\caption{
{
The exclusion {C.L.s} of all the UED models as functions of $M_\text{KK}$ obtained from the $S$ and $T$ parameters. 
Colors denote the same as in {Fig.}~\ref{fig:cutoff_upperbounds}. 
}
}
\label{fig:STplots}
\end{center}
\end{figure}

We also execute {a} $\chi^2$ {analysis} for putting indirect constraints on the UED models.
$\chi^2$ from $S$ and $T$ is defined as
\al{
\chi^2_{ST} = \frac{1}{1-\rho_{ST}^2} \left\{
\frac{\left( S - S_{\text{exp.}} \right)^2}{\sigma_S^2} +
\frac{\left( T - T_{\text{exp.}} \right)^2}{\sigma_T^2} -
\frac{2\rho_{ST}}{\sigma_S \sigma_T} \left( S - S_{\text{exp.}} \right) \left( T - T_{\text{exp.}} \right)
\right\},
}
where $S$ and $T$ are the theoretical inputs in Eqs.~(\ref{Sform}) and (\ref{Tform}) and the others are the experimental resultants in Eq.~(\ref{ST_experimental}).
{In this and the next sections, we again adopt the assumption of $R_5 = R_6 = R$.}

At first in this section, we consider the possibility without threshold correction to $S$ and $T$ in Eq.~(\ref{ST_threshold}).
We consider the maximal cutoffs with $M_{\text{KK}} = 1\,\text{TeV}$ irrespective of $M_{\text{KK}}$ because our interest is in the case that $M_{\text{KK}}$ is about a few TeV{,} and the values are almost universal as a function of $M_{\text{KK}}$ in each model around a few TeV as shown in Fig.~\ref{fig:cutoff_upperbounds}. 
The result is {listed} in Fig.~\ref{fig:STplots}.
{The plots for a wider range of $M_{\text{KK}}$ {are shown} in Appendix~\ref{Appendix:bounds},} where we can find the global minima in every curve.
Each minimum is located around
${1700}\,\text{GeV}$ ($T^2/Z_2$),
$1500\,\text{GeV}$ {($T^2/(Z_2 \times Z'_2)$)},
$1300\,\text{GeV}$ ($T^2/Z_4$),
$2200\,\text{GeV}$ ($S^2$),
$1500\,\text{GeV}$ ($S^2/Z_2$),
{$1800\,\text{GeV}$ ($RP^2$)},
$2000\,\text{GeV}$ ({PS}), and
$1000\,\text{GeV}$ (mUED).
Interestingly, these values are somewhat greater than the corresponding $95\%$ {C.L.} bound from the combined results in the Higgs searches as shown in Fig.~\ref{Fig:BoundHiggs}.

We also estimate the $95\%$ {C.L.} bounds of the models from Fig.~\ref{fig:STplots}, and the values are about 
$1200\,\text{GeV}$ ($T^2/Z_2$),
$1100\,\text{GeV}$ {($T^2/(Z_2 \times Z'_2)$)},
$900\,\text{GeV}$ ($T^2/Z_4$),
$1500\,\text{GeV}$ ($S^2$),
${1100}\,\text{GeV}$ ($S^2/Z_2$),
{$1200\,\text{GeV}$ ($RP^2$)},
$1400\,\text{GeV}$ ({PS}), and
${700}\,\text{GeV}$ (mUED).
Here{,} we can notice that these indirect bounds are compatible with the direct bounds via the LHC results discussed in the previous section. 
{We note} that our $95\%$ {C.L.} bound (${700}\,\text{GeV}$) on the mUED is close to the previous values by the Gfitter group ($700\,\text{GeV}$ in $m_H = 126\,\text{GeV}$) in Ref.~\cite{Baak:2011ze}.

\subsection{Numerical results with threshold correction}

\begin{figure}[t]
\begin{center}
   \includegraphics[width=0.99\columnwidth]{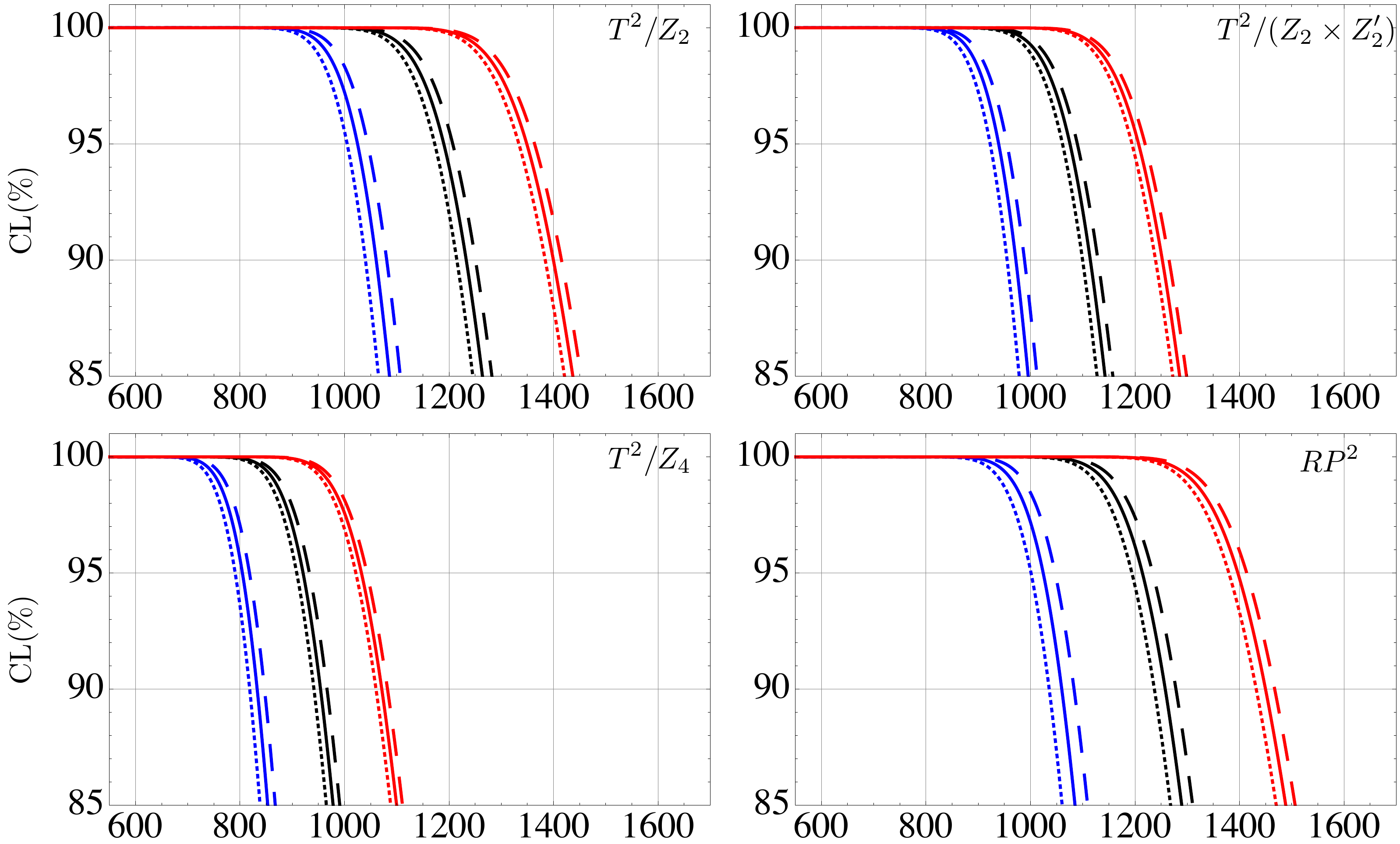}
   \includegraphics[width=0.99\columnwidth]{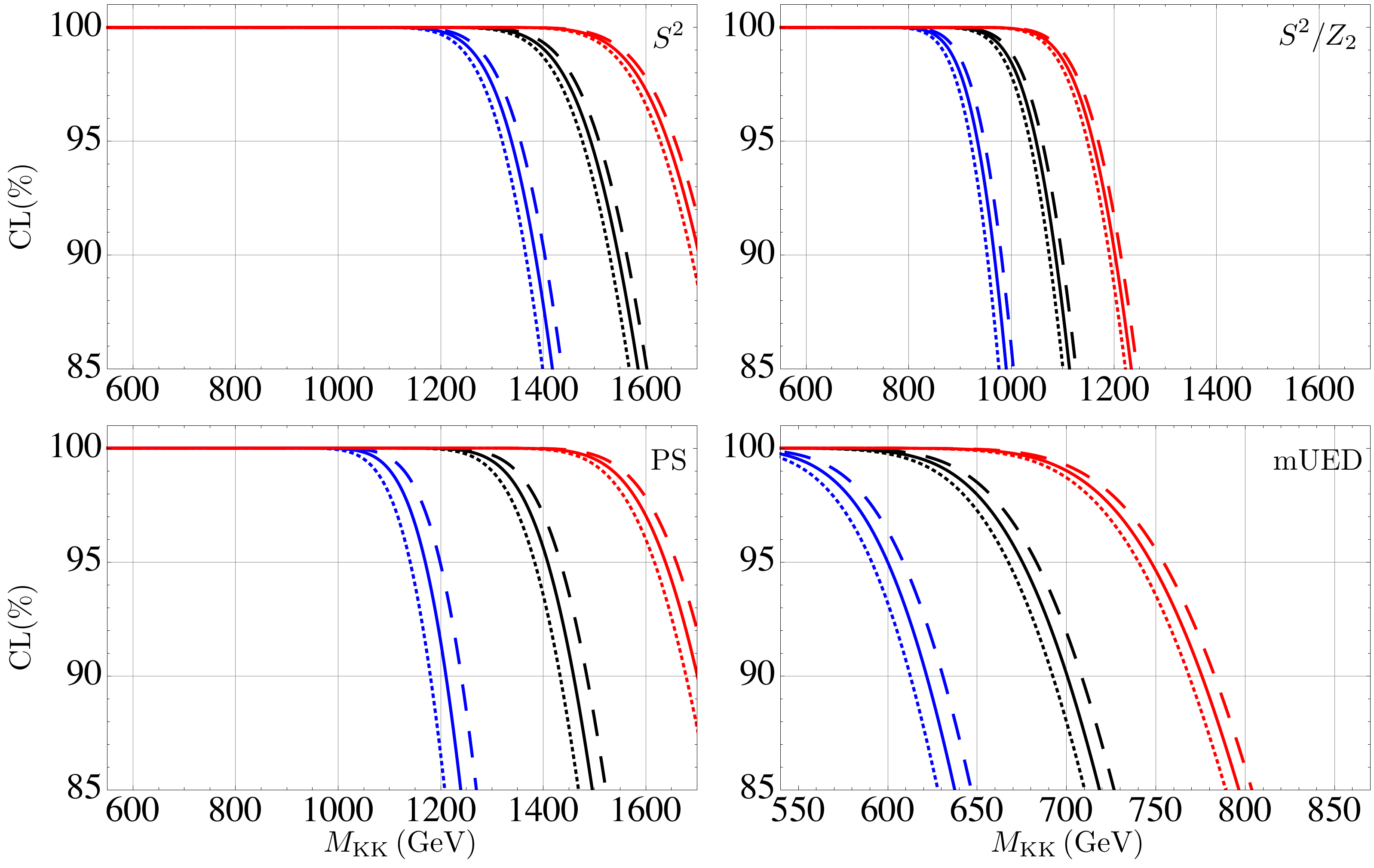}
\caption{
{The exclusion C.L.s via the $\chi^2$ analysis of $S$ and $T$ parameters as a function of $M_\text{KK}$ with the maximal cutoffs in the UED models in Table~\ref{table:maximalcutoff} with threshold corrections in Eq.~(30). In each panel corresponding to each model, the left (blue), center (black), and right (red) bunches of lines are for $c_S=+1$, 0, and $-1$, respectively. In each bunch, the dotted, solid, and dashed lines correspond to $c_T=-1$, 0, and $+1$, respectively.}
}
\label{fig:STplots2}
\end{center}
\end{figure} 

As we have seen in {Sec.}~\ref{section:2},
the vacuum is destabilized rapidly {for the Higgs mass} $m_H = 126\,\text{GeV}$, and we should take the cutoff scale quite a low.
For completeness, we estimate the threshold corrections via physics around the cutoff scale $\Lambda$.

In this section, 
{we summarize the results, for the maximum UV cutoff scale{,} given in Table~\ref{table:maximalcutoff}, in Fig.~\ref{fig:STplots2}.
}
Here{,} we examine the three extremal possibilities ($c_{S,T} = 0, +1, -1$) {for} each {of} the two {coefficients} $c_S$ and $c_T$ in Eq.~(\ref{ST_threshold}).

We can find sizable deviations from the case of $c_S = c_T = 0$ in all the {models}.
The corrections from $S_{\text{threshold}}$ are significant and the $95\%$ {C.L.} bounds turn out to be modified {by} the magnitudes about $100 \sim 200\,\text{GeV}$, depending on the models, toward both positive and negative directions,
while the corrections from $T_{\text{threshold}}$ are subleading. 
{We note} that we can find the global minimum in all the models after taking into account the threshold corrections, which is not shown in Fig.~\ref{fig:STplots2}.

One important thing is that{,} even in the 5D mUED, the threshold corrections {are more significant than was thought.}
We report that the $T^2/Z_2$ case was studied in Ref.~\cite{Appelquist:2002wb} with $\Lambda = 5 M_{\text{KK}}$.
In our result, the degree of the deviations from {the case} without {the} threshold correction is enlarged since we can take $\Lambda$ {at most} $2.5 M_{\text{KK}}$ as {shown} in Table~\ref{table:maximalcutoff}.

\section{Summary and Discussions
\label{section:summary}}

We have studied the effects from the KK particles in the UED models on the Higgs searches at the LHC and on the electroweak precision data. Both are dependent on the UV cutoff scale $\Lambda$ of the {higher-dimensional} theory. We have evaluated the highest possible $\Lambda$ consistent with the vacuum stability bound on the Higgs potential.

In the UED models, the contributions {from} loop diagrams including the KK top quarks and gauge bosons modify the Higgs decay rate and production cross section, 
{which} affect the Higgs signal strengths at the LHC. 
On the other hand, the KK excited states of the {heavier} SM particles (top quark, Higgs boson{,} and {the massive} gauge bosons) alter the $S$ and $T$ parameters.
From the analysis on the results of the Higgs signal strengths in the decay modes $H \to \gamma\gamma$, $ZZ, WW$ and of the $S$ and $T$ parameters, 
we have estimated the two types of bounds on the KK scales in 5D and 6D UED models, {which are} summarized in Table~\ref{Tab:boundsummary}. 
Comparing the former bounds with the latter, we find that the latter are slightly {more severe} than the former. 
However, in {few years} the Higgs searches at the LHC will put stronger constraints on the KK scale in the UED models.
We note that there remain uncertainties from the choices of the UV cutoff $\Lambda<\Lambda_\text{max}$, the {higher-dimensional} operators there, and the {low-energy} input for the top Yukawa coupling.
{In {the} estimation {of} $\Lambda_{\text{max}}$, we focus on the} {vacuum stability bound, namely{,} the condition~(\ref{cutoff_criterion}) on the coupling $\lambda$.
This new bound is tighter than the conventional one derived from the perturbativity of the gauge couplings.
It might be interesting to take into account the effects of {higher-dimensional} operators for the stability argument, as we have done for the $S$ and $T$, since the scale of $\Lambda_{\text{max}}$ tends to be low in the UED models after the Higgs discovery.
}

\begin{table}[t]
\begin{center}
\begin{tabular}{c|c|c|c}
\hline\hline
                                        & $\Lambda/M_\text{KK}$ for $M_\text{KK} \sim O(\text{TeV})$  & Higgs signal strength  & $S$ and $T$ parameters  \\ \hline
mUED                             & 5.0                                                                                                & {610} GeV                      & {680} GeV                 \\ \hline
$T^2/Z_2$                       & 2.5                                                                                                & {1060} GeV                   & 1190 GeV                 \\\hline
$T^2/(Z_2\times Z_2')$   & 2.9                                                                                                & {960} GeV                      & {1080} GeV                 \\ \hline
$T^2/Z_4$                       & 3.4                                                                                                & {820} GeV                      & {920} GeV                \\ \hline
$RP^2$                           & {2.3}                                                                                                & {1060}  GeV                   & {1220}   GeV              \\ \hline
$S^2$                              & 2.3                                                                                                & {1330} GeV                      & {1490} GeV                 \\\hline
$S^2/Z_2$                       & 3.2                                                                                                & {940} GeV                      & {1050} GeV                  \\ \hline
PS                                   & 1.9                                                                                                & {1240} GeV                     & 1410 GeV                   \\ 
\hline\hline
\end{tabular}
\caption{Highest possible UV cutoff scales and lower bounds on the KK scale $M_\text{KK}$ for {each model} at the 95\% {C.L.}} 
\label{Tab:boundsummary}
\end{center}
\end{table}

\begin{figure}
\begin{center}
   \hfill
   \includegraphics[viewport=0 0 955 325, width=32em]{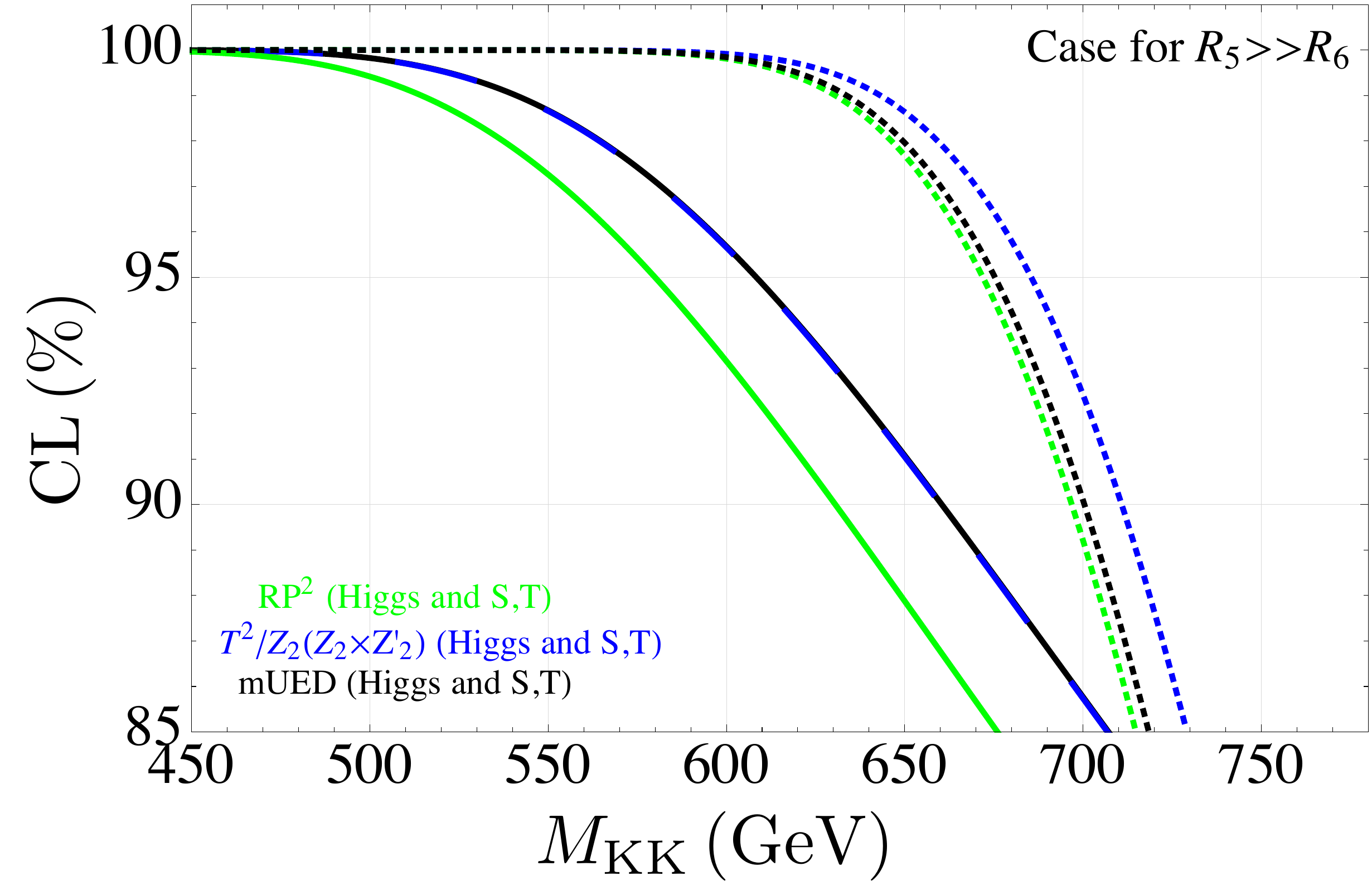} \hfill\\
\caption{
{
The same exclusion {C.L.s} as above, in the limit $R_5\gg R_6$.
The solid and dotted lines correspond to the bounds from Higgs searches and $S$, $T$ constraints, respectively. The former Higgs bounds for $T^2/Z_2(Z_2\times Z_2')$ and mUED are degenerate with each other.
}
}
\label{5D_limit}
\end{center}
\end{figure} 

{
The authors of Ref.~\cite{Arbey:2012ke} have shown that the case $R_5=R_6$ is disfavored in the $RP^2$ model if we identify the lowest KK state to be the dark matter, compared with the limit $R_5\gg R_6$, because the former requires the lower dark matter mass than the latter. We show in Fig.~\ref{5D_limit} the same exclusion limits from the Higgs searches and from $S$, $T$, as given above, in this five dimensional limit. As expected, the bounds become almost the same as that in the mUED model, expect for the small difference due to extra contributions from the 6D KK states of the gauge field. Note also that the five dimensional limit $R_5\gg R_6$ loosens the vacuum stability bound and hence allows the higher cutoff scale $\Lambda_\text{max}=5.5M_\text{KK}$ and $3.9M_\text{KK}$ for $T_2/Z_2(Z_2\times Z_2')$ and $RP^2$, respectively.
The higher the cutoff scale is, the larger the number of KK states below it. We have chosen $\Lambda=\Lambda_\text{max}$ and taken into account this effect in Fig.~\ref{5D_limit}.
}

We briefly comment on other bounds.
We can find the recent studies in {bounds} from collider simulations in the mUED, 6D UEDs on $T^2/Z_4${,} and $RP^2$.
\begin{itemize}
\item On mUED in Ref.~\cite{Belyaev:2012ai}:
$M_{\text{KK}} \gtrsim 1300\,\text{GeV}$ with $95\%$ {C.L. through trilepton} signature $+$Missing ET (MET) with $20\,\text{fb}^{-1}$ at $\sqrt{s} = 8\,\text{TeV}$ $(\Lambda R = 10)$.
\item On $T^2/Z_4$ in Ref~\cite{Choudhury:2011jk}:
$M_{\text{KK}} \gtrsim 500\,\text{GeV}$
with $5\sigma$ {C.L.} through $n$-jets$+\gamma+$MET ($n \geq 4$)
with $2\,\text{fb}^{-1}$ at $\sqrt{s} = 7\,\text{TeV}$.
\item On $RP^2$ in Ref.~\cite{Cacciapaglia:2013wha}:
$M_{\text{KK}} \gtrsim 600\,\text{GeV}$ with above $99\%$ {C.L.} through
CMS $\alpha_T$ analysis in leptons$+$MET with
$\sim5\,\text{fb}^{-1}$ at $\sqrt{s} = 7\,\text{TeV}$.
\end{itemize}
The constraint on the mUED is tighter than {ours} from the direct Higgs search and $S$, $T$ parameters, while these on $T^2/Z_4$ and $RP^2$ (with the limited integrated luminosities) {are} somewhat loose compared with ours.
As pointed out in Ref.~\cite{Datta:2012db}, a UED model with a low cutoff scale results in {a} much compressed KK spectrum and hence 
{becomes difficult to {detect} at the LHC.}
It is noted that such a degenerate possibility has not been explored enough{,} and the analysis with $M_{T2}$ and/or event shape variables is suitable for the case~\cite{Murayama:2011hj,Datta:2011vg}.
We also refer to the bounds from dark matter relic abundance in {these} three models.
The {upper} bound {on} $M_{\text{KK}}$ {is} {approximately} less than $200\,\text{GeV}$ in $T^2/Z_4$~\cite{Dobrescu:2007ec} and $470\,\text{GeV}$ in $RP^2$ with $R_5=R_6$~\cite{Arbey:2012ke}, to circumvent an overabundance of matter in the {Universe}.
This {bound} suggests that these 6D UEDs on both geometries are disfavored in {combination with} our results.\footnote{
{In {the} $RP^2$ case, the upper bound on $M_{\text{KK}}$ can be uplifted around $1.5\,\text{TeV}$ by introducing tuning in the parameter space~\cite{Cacciapaglia:2013wha,Arbey:2012ke}.}
}
In the mUED, the range being consistent with the relics is $1300\,\text{GeV} \lesssim M_{\text{KK}} \lesssim 1500\,\text{GeV}$~\cite{Belanger:2010yx}, which is just an unexplored area.

We have {studied} the suppression {effects} of Higgs decay into diphoton in Sec.~\ref{section:LHC}. 
These effects {can} also affect the measuring of the Higgs to diphoton coupling at a future linear collider~\cite{Kakuda:2012px}. 
{We summarize in Table~\ref{Tab:predictionsummary}} {the ratio of} $\BR(H \to \gamma\gamma)$ {as well as} the Higgs production {cross section} from $\gamma\gamma$ collision in each UED model to those in the SM 
{for the lowest possible KK scale with the highest possible UV cutoff.}
We find that the branching ratio and the Higgs production can be suppressed by a factor $\sim 0.9$ compared with SM. 
This is marginally accessible at the linear collider with integrated luminosity 500 fb$^{-1}$ at 500GeV {for which the} expected precision for the $\BR(H \to \gamma\gamma)$ is $23\%$ for $M_h =120$ GeV~\cite{Desch:2003xq}.
This precision is refined to 5.4\% with luminosity 1 ab$^{-1}$ at 1 TeV for the same Higgs mass~\cite{Heinemeyer:2005gs}. 
When we employ the {photon-photon} collider option, $H\gamma\gamma$ coupling can be measured more directly {from} the total production cross section of the Higgs.
This is well within the reach for an integrated photon-photon luminosity 410 fb$^{-1}$ at a linear $e^+ e^-$ collider operated at a $\sqrt{s} = 210$ GeV, 
{which} can measure $\Gamma_{H\rightarrow \gamma\gamma} \times\BR(H\rightarrow \gamma\gamma)$ with accuracy of 2.1$\%$ for $m_H=$ 120\,GeV~\cite{Barklow:2003hz}. 
\begin{table}[t]
\begin{center}
\begin{tabular}{c|c|c}
\hline\hline
                                        & UED/SM ratio of $\BR(H \to \gamma\gamma)$  & UED/SM ratio of $\sigma_{\gamma\gamma \to H}$   \\ \hline
mUED                             & 0.93                                                                     & 0.94                 \\ \hline
$T^2/Z_2$                       & 0.93                                                                     & 0.94                 \\ \hline
$T^2/(Z_2\times Z_2')$   & 0.93                                                                     & 0.94                 \\ \hline
$T^2/Z_4$                       & 0.92                                                                     & 0.94                  \\ \hline
$RP^2$                           &{ 0.93}                                                                     & {0.94}                  \\ \hline
$S^2$                              & 0.85                                                                     & 0.88                  \\ \hline
$S^2/Z_2$                       & 0.92                                                                     & 0.94                  \\ \hline
PS                                   & 0.90                                                                     & 0.92                   \\ 
\hline\hline
\end{tabular}
\caption{Prediction on the UED/SM ratio of $\BR(H \to \gamma\gamma)$ and $\sigma_{\gamma\gamma \to H}$ with the lowest {possible value} of the KK scale. } 
\label{Tab:predictionsummary}
\end{center}
\end{table}

\section*{Acknowledgement}
We thank Tomohiro Abe for useful comments on oblique corrections {and} 
Swarup Kumar Majee for discussions {in} the early stages of this work.
K.N.\ is grateful for valuable discussions with Joydeep Chakrabortty and Daisuke Harada and for fruitful conversations with Anindya Datta and Sreerup Raychaudhuri.
K.N.\ is partially supported by funding available from the Department of 
Atomic Energy, Government of India for the Regional Centre for Accelerator-based
Particle Physics (RECAPP), Harish-Chandra Research Institute.
The work of K.O. {(R.W.)} is in part supported by {the Grands-in-Aid for Scientific Research No.~23104009, No.~20244028, and No.~23740192 {(No.~248920)}.}


\appendix
\section*{Appendix}

\section{RGEs in 6D UED models and mUED \label{Appendix_A}}
In this {appendix}, we show the concrete forms of RGEs for gauge, Yukawa, and Higgs {self-couplings}.
Here{,} we rewrite the schematic shape of the beta function for a quantity $\Q$ in 6D UED:
\al{
\beta_\Q = \beta_\Q^{\text{(SM)}} + \sum_{s:\,\text{KK states}} \theta(\mu - M_s) \Big( N_s
\beta^{\text{(KK)}}_{s,\Q} \Big).
}
Details of this expression are found in {Sec.}~\ref{Section:2-1}.
As we have already discussed {there}, the beta functions take different forms {depending on} the following two categories: UEDs on {an orientable space} and {those} on {an unorientable one}. 
The former contains 
$T^2/Z_2$, {$T^2/(Z_2 \times Z_2')$}, $T^2/Z_4$, $S^2$, and $S^2/Z_2$ and
the latter the remains $RP^2$, {PS}.
{The RGEs obtained in this work are consistent with those obtained for mUED~\cite{Bhattacharyya:2006ym} and for the SM~\cite{Cheng:1973nv}.}
The contribution of the KK particles to the beta function $\beta^{\text{(KK)}}_{s,\Q}$ is independent of the KK index and we can omit the index $s$ as $\beta^{\text{(KK)}}_{\Q}$.
We already explained the reason in {Sec.}~\ref{Section:2-1}.
{We note that in all the RGE analyses in this paper, we ignore Yukawa couplings except for {the} top quark one.}

\subsection{UEDs {in orientable space}}

{In the following,} $\lambda$ is 4D Higgs {self-couplings}; $g_1$, $g_2$, and $g_3$ show the 4D $U(1)_Y$, $SU(2)_W$, and $SU(3)_C$ gauge couplings;
$y_{\ell_k}, y_{u_k}, y_{d_k}\,(k=1,2,3)$ represent the 4D (diagonalized) Yukawa couplings of the charged leptons, the up-type quarks, and the down-type quarks, respectively.
Here{,} we adopt the SM normalization in the $U(1)_Y$ gauge coupling $g_1$.
The index $k$ indicates their generations.
{$V_{ij}$ means the Cabibbo--Kobayashi--Maskawa matrix{,} and $N_{C_{f_i}}$ indicates the color factor of the particle $f_i$, namely{,} $3$ for quarks and $1$ for leptons.}

\medskip

\underline{$\circ\ \Q = \lambda$}
\al{
\beta^{(\text{SM})}_{\lambda} &= \frac{1}{(4\pi)^2} \Bigg\{ 6\lambda^2 - (3g_1^2 + 9g_2^2) \lambda + \frac{3}{2} (g_1^4 + 2g_1^2 g_2^2 + 3g_2^4) + 4\lambda \sum_{i} N_{C_{f_i}} y_{f_i}^2 -8 \sum_{i} N_{C_{f_i}} y_{f_i}^4 \bigg\}, \\
\beta^{(\text{KK})}_{\lambda} &= \frac{1}{(4\pi)^2} \Bigg\{  6\lambda^2 - (3g_1^2 + 9g_2^2) \lambda + \left(\frac{5}{2} g_1^4 + 5 g_1^2 g_2^2 + \frac{15}{2} g_2^4 \right) + 8\lambda \sum_{i} N_{C_{f_i}} y_{f_i}^2 - 16 \sum_{i} N_{C_{f_i}} y_{f_i}^4 \Bigg\}.
}

\medskip

\underline{$\circ\ \Q = g_i\,(i=1,2,3)$}
\al{
\beta^{\text{(SM)}}_{g_i}     = \frac{1}{(4\pi)^2} b^{\text{(SM)}}_{g_i} g_i^3,\quad
\beta^{\text{(KK)}}_{g_i}     = \frac{1}{(4\pi)^2} b_{g_i}^{\text{(KK)}} g_i^3,
}
with $b^{\text{(SM)}}_{g_i} = (\frac{41}{6}, -\frac{19}{6}, -7)$ and $b^{\text{(KK)}}_{g_i} = (\frac{27}{2}, \frac32, -2)$ 
for $g_i = (g_1,g_2,g_3)$, respectively.

\medskip

\underline{$\circ\ \Q = y_{\ell_k}, y_{u_k}, y_{d_k}\,(k=1,2,3)$}
\al{
\beta^{\text{(SM)}}_{y_{\ell_k}}    &= \frac{1}{(4\pi)^2} \Bigg\{ -\frac{15}{4} g_1^2 -\frac94 g_2^2 + \frac32 {y_{\ell_k}^2} + \sum_{i} N_{C_{f_i}} y_{f_i}^2 \Bigg\} {y_{\ell_k}}, \\
\beta^{\text{(SM)}}_{y_{u_k}}    &= \frac{1}{(4\pi)^2} \Bigg\{ -\frac{17}{12} g_1^2 -\frac94 g_2^2 -8g_s^2 + \frac32 {y_{u_k}^2} + \sum_{j} y_{d_j}^2 {(V_{kj} V^{\dagger}_{jk})} + \sum_{i} N_{C_{f_i}} y_{f_i}^2 \Bigg\} y_{u_k}, \\ 
\beta^{\text{(SM)}}_{y_{d_k}}    &= \frac{1}{(4\pi)^2} \Bigg\{ -\frac{15}{12} g_1^2 -\frac94 g_2^2 -8g_s^2 + \frac32 {y_{d_k}^2} + \sum_{j} y_{u_j}^2 {(V^{\dagger}_{kj} V_{jk})} + \sum_{i} N_{C_{f_i}} y_{f_i}^2 \Bigg\} y_{d_k}, \\
\beta_{y_{\ell_k}}^{(\text{KK})} &= \frac{1}{(4\pi)^2} \Bigg\{
		- \frac{9}{2} g_1^2 - \frac{3}{2} g_2^2 + \frac{3}{2} {y_{\ell_k}^2}
		+ 2 \sum_i N_{C_{f_i}} y_{f_i}^2 \Bigg\} y_{\ell_k}, \\
\beta_{y_{u_k}}^{(\text{KK})} &= \frac{1}{(4\pi)^2} \Bigg\{
		- \frac{25}{18} g_1^2 - \frac{3}{2} g_2^2 - \frac{32}{3} g_s^2 + \frac{3}{2} {y_{u_k}^2}
		+ 2 \sum_i N_{C_{f_i}} y_{f_i}^2   -\frac{3}{2} { \sum_{j} (V_{kj} V_{jk}^{\dagger})} y_{d_j}^2 \Bigg\} y_{u_k}, \\
\beta_{y_{d_k}}^{(\text{KK})} &= \frac{1}{(4\pi)^2} \Bigg\{
		- \frac{1}{18} g_1^2 - \frac{3}{2} g_2^2 - \frac{32}{3} g_s^2 + \frac{3}{2} {y_{d_k}^2}
		+ 2 \sum_i N_{C_{f_i}} y_{f_i}^2   -\frac{3}{2} { \sum_{j} (V_{kj}^{\dagger} V_{jk})} y_{u_j}^2 \Bigg\} y_{d_k}.
}

\subsection{{PS} case}

In the case of {PS}, the contributions of the bosonic KK particles to the beta functions {is} classified into two {categories} as $\beta^{\text{(KK)}}_{\text{even},\Q}$ and $\beta^{\text{(KK)}}_{\text{odd},\Q}$.

\medskip

\underline{$\circ\ \Q = \lambda$}
\al{
\beta_{\text{even}, \lambda}^{(\text{KK})} &= \frac{1}{(4\pi)^2} \Bigg\{
		6 \lambda^2 -3 \lambda g_1^2 -9 \lambda g_2^2
		+ 2 g_1^4 + 4 g_1^2 g_2^2 + 6 g_2^4 
		+ 8 \sum_i \lambda N_{C_{f_i}} y_{f_i}^2 -16 \sum_i N_{C_{f_i}} y_{f_i}^4 \Bigg\}, \\
\beta_{\text{odd}, \lambda}^{(\text{KK})} &= \frac{1}{(4\pi)^2} \Bigg\{
		+ \frac{1}{2} g_1^4 + g_1^2 g_2^2 + \frac{3}{2} g_2^4 
		+ 8 \sum_i \lambda N_{C_{f_i}} y_{f_i}^2 -16 \sum_i N_{C_{f_i}} y_{f_i}^4 \Bigg\}.
}

\medskip

\underline{$\circ\ \Q = g_i\,(i=1,2,3)$}


\al{
\beta^{\text{(KK)}}_{\text{even}, g_i} = \frac{1}{(4\pi)^2} b^{\text{(KK)}}_{\text{even}, g_i} g_i^3,\quad
\beta^{\text{(KK)}}_{\text{odd}, g_i} = \frac{1}{(4\pi)^2} b_{\text{odd}, g_i}^{\text{(KK)}} g_i^3,
}
with $b^{\text{(KK)}}_{\text{even}, g_i} = (\frac{27}{2}, \frac{7}{6}, -\frac{5}{2})$ and $b^{\text{(KK)}}_{\text{odd}, g_i} = (\frac{40}{3}, \frac{25}{3}, \frac{17}{2})$ 
for $g_i = (g_1,g_2,g_3)$, respectively.

\medskip

\underline{$\circ\ \Q = y_{\ell_k}, y_{u_k}, y_{d_k}\,(k=1,2,3)$}
\al{
\beta_{\text{even}, y_{\ell_k}}^{(\text{KK})} &= \frac{1}{(4\pi)^2} \Bigg\{
		- \frac{33}{8} g_1^2 - \frac{15}{8} g_2^2 + \frac{3}{2} {y_{\ell_k}^2}
		+ 2 \sum_i N_{C_{f_i}} y_{f_i}^2 \Bigg\} y_{\ell_k}, \\
\beta_{\text{even}, y_{u_k}}^{(\text{KK})} &= \frac{1}{(4\pi)^2} \Bigg\{
		- \frac{101}{72} g_1^2 - \frac{15}{8} g_2^2 - \frac{28}{3} g_s^2 + \frac{3}{2} {y_{u_k}^2}
		+ 2 \sum_i N_{C_{f_i}} y_{f_i}^2   -\frac{3}{2} { \sum_{j} (V_{kj} V_{jk}^{\dagger})} y_{d_j}^2 \Bigg\} y_{u_k}, \\
\beta_{\text{even}, y_{d_k}}^{(\text{KK})} &= \frac{1}{(4\pi)^2} \Bigg\{
		- \frac{17}{72} g_1^2 - \frac{15}{8} g_2^2 - \frac{28}{3} g_s^2 + \frac{3}{2} {y_{d_k}^2}
		+ 2 \sum_i N_{C_{f_i}} y_{f_i}^2   -\frac{3}{2} { \sum_{{j}} (V_{kj}^{\dagger} V_{jk})} y_{u_j}^2 \Bigg\} y_{d_k}, \\
\beta_{\text{odd}, y_{\ell_k}}^{(\text{KK})} &= \frac{1}{(4\pi)^2} \Bigg\{
		- \frac{3}{8} g_1^2 + \frac{3}{8} g_2^2
		+ 2 \sum_i N_{C_{f_i}} y_{f_i}^2 \Bigg\} y_{\ell_k}, \\
\beta_{\text{odd}, y_{u_k}}^{(\text{KK})} &= \frac{1}{(4\pi)^2} \Bigg\{
		+ \frac{1}{72} g_1^2 + \frac{3}{8} g_2^2 - \frac{4}{3} g_s^2
		+ 2 \sum_i N_{C_{f_i}} y_{f_i}^2 \Bigg\} y_{u_k}, \\
\beta_{\text{odd}, y_{d_k}}^{(\text{KK})} &= \frac{1}{(4\pi)^2} \Bigg\{
		+ \frac{13}{72} g_1^2 + \frac{3}{8} g_2^2 - \frac{4}{3} g_s^2
		+ 2 \sum_i N_{C_{f_i}} y_{f_i}^2 \Bigg\} y_{d_k}.
}

\subsection{$RP^2$ case}


In the regions {having bosonic modes} ({regions I, II, and III}), the following relations are fulfilled for {each type of coupling} $C$:
\al{
\beta_{\text{region I}, C}^{(\text{KK})} = \beta_{\text{even}, C}^{(\text{KK})}, \quad
\beta_{\text{region II}, C}^{(\text{KK})} =  \beta_{\text{odd}, C}^{(\text{KK})}, \quad
\beta_{\text{region III}, C}^{(\text{KK})} = \beta_{C}^{(\text{KK})}.
}
We write down {the formula for} the region without {having a bosonic mode} (region IV).

\medskip

\underline{$\circ\ \Q = \lambda$}
\al{
\beta_{\text{region IV}, \lambda}^{(\text{KK})} &= \frac{1}{(4\pi)^2} \Bigg\{
		+ 8 \sum_i \lambda N_{C_{f_i}} y_{f_i}^2 -16 \sum_i N_{C_{f_i}} y_{f_i}^4 \Bigg\}.
}

\medskip

\underline{$\circ\ \Q = g_i\,(i=1,2,3)$}


\al{
\beta^{\text{(KK)}}_{\text{region IV}, g_i}     &= \frac{1}{(4\pi)^2} b^{\text{(KK)}}_{\text{region IV}, g_i} g_i^3,
}
with $b^{\text{(KK)}}_{\text{region IV}, g_i} = (\frac{40}{3}, 8, 8)$
for $g_i = (g_1,g_2,g_3)$, respectively.

\medskip

\underline{$\circ\ \Q = y_{\ell_k}, y_{u_k}, y_{d_k}\,(k=1,2,3)$}
\al{
\beta_{\text{region IV}, y_{\ell_k}}^{(\text{KK})} &= \frac{1}{(4\pi)^2} \Bigg\{
		+ 2 \sum_i N_{C_{f_i}} y_{f_i}^2 \Bigg\} y_{\ell_k}, \\
\beta_{\text{region IV}, y_{u_k}}^{(\text{KK})} &= \frac{1}{(4\pi)^2} \Bigg\{
		+ 2 \sum_i N_{C_{f_i}} y_{f_i}^2 \Bigg\} y_{u_k}, \\
\beta_{\text{region IV}, y_{d_k}}^{(\text{KK})} &= \frac{1}{(4\pi)^2} \Bigg\{
		+ 2 \sum_i N_{C_{f_i}} y_{f_i}^2 \Bigg\} y_{d_k}.
}

\subsection{mUED case}

{The surviving modes for each} KK level in the mUED {are} totally the same {as in} region I of the $RP^2$ or {in the} ``even" region of the {PS}.
Hence{,} we can use {those} forms for RGEs in the mUED.
We can check that our results of this part are consistent with those in Refs.~\cite{Bhattacharyya:2006ym,Cornell:2011ge}.

\section{Loop functions in single Higgs production and decay
\label{Appendix:loopfunction}}

In this {appendix}, we summarize the loop functions {that} are needed for estimating the single Higgs production through the gluon fusion process and the Higgs decay into a pair of photons.
Readers who want more explanations on the above expressions {should} {consult} Ref.~\cite{Nishiwaki:2011gm}.

For each model, the loop function $J_t^{\text{model}}$ describes the contributions of all the zero and KK modes for the top quark in the triangle loops:
\al{
 J_t^\text{SM} (\hat s)
 	&=  I\fn{ m_t^2 \over \hat s }, \label{result_of_SM} \\
 J_t^\text{mUED} (\hat s)
 	&=  \br{ I\fn{ m_t^2 \over \hat s } +2 \sum_{n\geq1} \left( {m_t \over m_{t(n)}} \right )^2 I\fn{ {m_{t(n)}^2 \over \hat s} } },\label{result_of_mUED}  \\
	{J_t^{T^2/Z_2}(\hat s) = J_t^{{RP}^2}(\hat s)}
		&=  \br{ I\fn{ m_t^2 \over \hat s } +2 \sum_{{m+n \geq 1 \atop \text{or\ } m=-n \geq 1}}  \left( {m_t \over m_{t(m,n)}} \right )^2 I\fn{ m_{t(m,n)}^2 \over \hat s } },
	\label{result_of_T2Z2}\\
 {J_t^{T^2/Z_4}(\hat s)}
 	&=  \br{ I\fn{ m_t^2 \over \hat s } +2 \sum_{m\geq1, n\geq0} \left( {m_t \over m_{t(m,n)}} \right )^2 I\fn{ m_{t({m,n})}^2 \over \hat s } },
 \label{result_of_T2Z4}\\ 
	{J_t^{T^2/(Z_2\times Z'_2)}(\hat s)}
		&=	 \br{ I\fn{ m_t^2 \over \hat s } +2 \sum_{m\geq0,n \geq 0, \atop (m,n) \not= (0,0)}  \left( {m_t \over m_{t(m,n)}} \right )^2 I\paren{ m_{t(m,n)}^2 \over \hat s } },
	\label{result_of_T2Z2Z2} \\
 J_t^{S^2/Z_2}(\hat s)
 	&=  \br{ I\fn{ m_t^2 \over \hat s } +2 \sum_{j\geq1}\left( {m_t \over m_{t(j)}} \right )^2 n^{S^2/Z_2}(j)\,I\paren{ m_{t(j)}^2 \over \hat s } },
 \label{result_of_S2Z2}\\
 J_t^\text{PS}(\hat s) = {J_t^{S^2}(\hat s)}
 	&=  \br{ I\fn{ m_t^2 \over \hat s } +2 \sum_{j\geq1}\left( {m_t \over m_{t(j)}} \right )^2 (2j+1)\,I\fn{ m_{t(j)}^2 \over \hat s } },
 \label{result_of_PS}
}
where 
$I$ is given by
\al{
I(\lambda)
	&=	{-2\lambda+\lambda(1-4\lambda)}
		\int_0^1{dx\over x}\ln\sqbr{{x(x-1)\over\lambda}+1-i\epsilon}.
}
{The} explicit result of the integral is
\al{
\int_0^1{dx\over x}\ln\sqbr{{x(x-1)\over\lambda}+1-i\epsilon}
	&=	\begin{cases}
		\displaystyle -2\sqbr{\arcsin{1\over\sqrt{4\lambda}}}^2
			&	\text{(for $\lambda\geq{1\over4}$)},\\
		\displaystyle {1\over2}\sqbr{
			\ln{1+\sqrt{1-4\lambda}\over1-\sqrt{1-4\lambda}}
			-i\pi
			}^2
			&	\text{(for $\lambda<{1\over4}$)},
		\end{cases}
}
where this form is related with the Passarino--Veltman's three-point scalar function $C_0$~\cite{Passarino:1978jh}.
{
$n^\text{model}(j)$ counts the number of degeneracy{,} and the explicit forms are {shown} in Eqs.~(\ref{S2count})--(\ref{PScount})
and we write the KK top and $W$ masses ($X=t,W$)} 
\begin{align}
 m_{X(n)}
 	&\equiv \sqrt{m_{X}^2 + {\frac{n^2}{R^2}}}, \\
 m_{X(m,n)}
 	&\equiv {\sqrt{m_{X}^2 + \frac{m^2}{R_5^2} + \frac{n^2}{R_6^2}}}, \\ 
 m_{X(j)}
 	&\equiv \sqrt{m_{X}^2 + {\frac{j(j+1)}{R^2}}}.
\end{align}
The range of the KK summation reflects the structure of each {extra-dimensional} background.\footnote{
The origin of the factor 2 in front of each KK summation is the fact {that} there are both {left- and right-handed} (namely, {vectorlike}) KK modes for each chiral quark zero mode corresponding to a SM quark.}
The loop functions which are needed for the process $H \rightarrow \gamma \gamma$ are as follows:
\begin{align}
 J_W^\text{SM} (m_H^2)
 	&=  L\fn{ {1\over2},3,3,6,0;{m_W^2\over m_H^2} ,{m_W^2\over m_H^2} }, \\
 J_W^\text{mUED} (m_H^2)
 	&=  J_W^\text{SM} (m_H^2)+ \sum_{n\geq1} L\fn{{1\over2},4,4,8,1;{m_W^2\over m_H^2},{{m_{W(n)}^2} \over m_H^2} }, \\
 J_W^{T^2/Z_4}(m_H^2)
 	&=  J_W^\text{SM} (m_H^2)+ \sum_{m\geq 1,n\geq 0} L\fn{ {1\over2},5,4,10,1;{m_W^2\over m_H^2},{{m_{W(m,n)}^2} \over m_H^2} },\label{T2Z4_W}\\ 
 J_W^{T^2/(Z_2\times Z'_2)}(m_H^2)
	&=  J_W^\text{SM} (m_H^2)+ \sum_{m\geq 0,n\geq 0\atop (m,n)\neq(0,0)} L\fn{ {1\over2},5,4,10,1;{m_W^2\over m_H^2},{{m_{W(m,n)}^2} \over m_H^2} },\\ 
 J_W^{T^2/Z_2}(m_H^2)
 	&=  J_W^\text{SM} (m_H^2)+ \sum_{{m+n \geq 1 \atop \text{or\ } m=-n \geq 1}} L\fn{ {1\over2},5,4,10,1;{m_W^2\over m_H^2},{{m_{W(m,n)}^2} \over m_H^2} },  \\
 J_W^{{RP}^2}(m_H^2)
	&=  J_W^\text{SM} (m_H^2)+ \sum_{(m,n)}^{A} L\fn{ {1\over2},4,4,8,1;{m_W^2\over m_H^2},{{m_{W(m,n)}^2} \over m_H^2} }
		+ \sum_{(m,n)}^{B} L\fn{ 0,1,0,2,0;{m_W^2\over m_H^2},{{m_{W(m,n)}^2} \over m_H^2} }, \\
 J_W^{S^2/Z_2}(m_H^2)
 	&=J_W^\text{SM} (m_H^2)+ \sum_{j\geq 1} n^{S^2/Z_2}(j)\,L\fn{ {1\over2},5,4,10,1;{m_W^2\over m_H^2},{{m_{W(j)}^2} \over m_H^2} },\\
 J_W^{S^2}(m_H^2)
 	&=J_W^\text{SM} (m_H^2)+\sum_{j\geq 1} (2j+1)\,L\fn{ {1\over2},5,4,10,1;{m_W^2\over m_H^2},{{m_{W(j)}^2} \over m_H^2} },\\
 J_W^\text{PS}(m_H^2)
  	&=	J_W^\text{SM} (m_H^2)+ \sum_{j\geq 1} \Bigg [ n_\text{even}^\text{PS}(j)\,L\fn{ {1\over2},4,4,8,1;{m_W^2\over m_H^2},{{m_{W(j)}^2} \over m_H^2} } \nn
	&\phantom{=	J_W^\text{SM} (m_H^2)+ \sum_{j\geq 1} \Bigg [}
		+n_\text{odd}^\text{PS}(j)\,L\fn{ 0,1,0,2,0;{m_W^2\over m_H^2},{{m_{W(j)}^2} \over m_H^2} } \Bigg ],\label{PS_W}
\end{align}
with
\begin{align}
L(a,b,c,d,e;\lambda_1,\lambda_2)
 &=	a+b\lambda_1- \left[\lambda_1 \fn{c-d\lambda_2} -e\lambda_2 \right ] \int_0^1{dx\over x}\ln\sqbr{{x(x-1)\over \lambda_2 }+1-i\epsilon}.
\end{align}
The {$A$ summation} for $RP^2$ {is} over the region that satisfies both $m\geq1$ and $n\geq1$ as well as over the ranges $(m,n)=(0,2), (0,4), (0,6), \dots$ and $(m,n)=(2,0), (4,0), (6,0), \dots$. Similarly, the {$B$ summation is} over $m\geq1$ and $n\geq1$ as well as over $(m,n)=(0,1), (0,3), (0,5), \dots$ and $(m,n)=(1,0), (3,0), (5,0), \dots$.

\section{{Two-point} functions of gauge bosons in 6D UEDs
\label{Appendix:twopoint}}

In this section, we summarize the {two-point} functions of photon, W{,} and Z bosons for calculating Peskin--Takeuchi {$S$ and $T$} parameters.

\subsection{Notations}
First{,} we summarize our notations for the Passarino--Veltman B function~\cite{Passarino:1978jh}.
In this section, we use the following descriptions for masses.
The mass squared of the {``$s$th"} KK mode of the particle $X$ is represented as
\al{
M_{X_s}^2 = m_X^2 + M_s^2
}
where $m_X$ is the corresponding zero-mode mass{,} and $M_s$ is the {$s$th} level KK mass.
Since only $Z$, $W$, $H${,} and top masses are not negligible compared with the KK scale $M_s$,
we use the representations
\al{
 M_{W_s}^2 &{:=} m_W^2 + M_s^2, \hspace{1em} M_{Z_s}^2 {:=} m_Z^2 + M_s^2 \notag \\
 M_{t_s}^2 &{:=} m_t^2 + M_s^2, \hspace{1em} M_{H_s}^2 {:=} m_H^2 + M_s^2 ,
}
and for the other fields,
\al{
M_{X_s}^2 {\simeq} M_s^2.
}

We will use the Passarino--Veltman loop integral to calculate {two-point} functions of the gauge bosons with external momentum $k$ below{,} and
the definition is
\al{
 \frac{1}{(4\pi)^2} B_{X_s,Y_s} (k^2) &= \int\frac{d^d p}{(2\pi)^d} \frac{1}{(p^2 - M_{X_s}^2)((p+k)^2 - M_{Y_s}^2)}
 \notag \\
		&= \frac{i}{(4\pi)^2} \left\{ \frac{1}{\overline{\varepsilon}}
		-\int^1_0 dx \ln \left [ (1-x) M_{X_s}^2 +x M_{Y_s}^2 -x(1-x)k^2 -i\epsilon \right] \right\},
}
where we use the dimensional regularization in $d$ dimensions and $\epsilon$ is an infinitesimal positive value.
$1/\overline{\varepsilon} \ (:= 1/\varepsilon - \gamma + \ln{4\pi})$ means the usual common divergent part with $\varepsilon = 2 - d/2$ and the Euler--Mascheroni constant $\gamma$.
The following short-hand description is also used later for simplicity{:}
\al{
B_{X_s} (k^2) := B_{X_s,X_s} (k^2),\quad
\delta B_{X_s,Y_s}(k^2) := { B_{X_s,Y_s}(k^2) -B_{X_s,Y_s}(0) \over k^2}.
}

Here{,} we write down some useful {formulas} for calculations:
\al{
B_{X_s,Y_s}(0) &\simeq \frac{1}{\overline{\varepsilon}} - \frac{1}{2} \frac{m_X^2 + m_Y^2}{M_s^2},
\\
B'_{X_s,Y_s}(0) &\simeq \frac{1}{6 M_s^2}, \\
B''_{X_s,Y_s}(0) &\simeq \frac{2}{3} \frac{1}{(m_X^2 - m_Y^2)^2} \frac{m_X^2 + m_Y^2}{M_s^2},
}
where we assume the hierarchy $m_X^2,\, m_Y^2 \ll M_s^2$ and values with a prime {mean} that it is differentiated with respect to $k^2$ once.

\subsection{Bosonic contributions to {the} two-point function of gauge bosons in 6D UEDs abd mUED}

In this section, we make a summary of bosonic two-point contributions to two-point function of gauge bosons in the 6D UEDs and the mUED for evaluating {$S$ and $T$} parameters.
For contributions of fermions, we can use the result in Ref.~\cite{Appelquist:2002wb}.

The generic form of a gauge boson {two-point} function is as follows:
\al{
\Pi^{\mu\nu}_{ab} (k^2) =  i \Pi_{ab}^{\text{T}} (k^2) \left( g^{\mu\nu} - \frac{k^{\mu}k^{\nu}}{k^2} \right) + i\Pi^{\text{L}}_{ab}(k^2) \frac{k^{\mu}k^{\nu}}{k^2} \label{4-1},
}
where $a$ and $b$ show {the} type of gauge bosons{,} and the superscript $\text{T}$ ($\text{L}$) indicates {the} transverse (longitudinal), respectively.

For estimating the {$S$ and $T$} parameters, we calculate only the transverse ones.
{In {each following subsection}, we show the contributions of KK bosonic particles to the {two-point} functions from the level-$s$ KK states.}

\subsubsection{UEDs on oriented geometry case}

\al{
 \Pi_{\gamma\gamma}^{\text{T},s}(k^2)
 &={\alpha \over 4 \pi} \Bigg\{ -{4\over9} k^2 +\left( \frac{7}{3} k^2 + \frac{20}{3} M_{W_s}^2 \right) B_{W_s}(k^2) - \frac{20}{3} M_{W_s}^2 B_{{W_s}}(0) \Bigg\},\\
 \Pi_{Z\gamma}^{\text{T},s}(k^2)
 &= {\alpha \over 4 \pi s_W c_W} \Bigg\{ \left( -\frac{1}{9} + \frac{4}{9} c_W^2 \right) k^2
    + \left( \frac{20}{3} c_W^2 - \frac{2}{3} \right) M_{W_s}^2 B_{W_s}(0) \notag \\ 
 & \quad + \left[ \left( -\frac{1}{6} - \frac{7}{3} c_W^2 \right) k^2 - \left( \frac{20}{3} c_W^2 - \frac{2}{3} \right) M_{W_s}^2
    -2 m_W^2 \right] B_{W_s}(k^2) \Bigg\}, \\
 \Pi_{ZZ}^{\text{T},s}(k^2)
 &={\alpha \over 4\pi s_W^2 c_W^2} \Bigg\{ 
    \left( \frac{2}{9} c_W^2 - \frac{4}{9} c_W^4 - {\frac{1}{18}} \right) k^2 \notag \\
 & \quad +\left( -\frac{20}{3} c_W^4 + \frac{4}{3} c_W^2 - \frac{1}{3} \right) M_{W_s}^2 B_{W_s}(0)
  - \frac{1}{6} M_{H_s}^2 B_{H_s}(0) - \frac{1}{6} M_{Z_s}^2 B_{Z_s}(0) \notag \\ 
 & \quad  +\left[ \left( \frac{7}{3} c_W^4 + \frac{1}{3} c_W^2 - \frac{1}{12} \right) k^2
   + \left( \frac{20}{3} c_W^4 - \frac{4}{3} c_W^2 + \frac{1}{3} \right) M_{W_s}^2
   - \left( 2 - 4 c_W^2 \right) m_W^2 \right] B_{{W_s}}(k^2) \notag \\
 & \quad  + \left( - \frac{1}{12} k^2 + \frac{1}{6} M_{Z_s}^2 + \frac{1}{6} M_{H_s}^2 
   - \frac{m_Z^2}{c_W^2} \right) B_{H_s,Z_s}(k^2)
   - \frac{1}{12} (M_{H_s}^2 - M_{{Z_s}}^2)^2 \delta B_{H_s, Z_s}(k^2) \Bigg \}, \\
 \Pi_{WW}^{\text{T},s}(k^2)
 &= {\alpha \over 4\pi s_W^2} \Bigg\{ 
   - \frac{1}{3} k^2  
   - \frac{1}{6} M_{H_s}^2 B_{H_s}(0) - 3 M_{W_s}^2 B_{W_s}(0) - \frac{17}{6} M_{Z_s}^2 B_{Z_s}(0)
   \notag \\ 
 & \quad  +\left[ -\frac{1}{12} k^2 +\frac{1}{6} M_{H_s}^2 +\frac{1}{6} M_{W_s}^2 - m_W^2 \right] B_{H_s,W_s}(k^2) \notag \\
 & \quad  +\left[ \frac{31}{12} k^2 +\frac{23}{6} \left( M_{W_s}^2 + M_{Z_s}^2 \right) 
   - 2 M_s^2 - m_Z^2 \left( c_W^2 -2 + c_W^{-2} \right) \right] B_{W_s,Z_s}(k^2) \notag \\
 & \quad  - \frac{1}{12} (M_{H_s}^2 - M_{W_s}^2)^2 \delta B_{H_s, W_s}(k^2) 
   - \frac{17}{12} (M_{Z_s}^2 - M_{W_s}^2)^2 \delta B_{Z_s, W_s}(k^2) \Bigg \},
}

\subsubsection{$RP^2$, $PS$, mUED cases}

As we have discussed in {Sec.}~\ref{section:2}, the particle contents of the region III of the $RP^2$ model is completely the same as those of the 6D UEDs on oriented geometries just as above{,} and we need not discuss them.
Based on the knowledge in {Sec.}~\ref{section:2}, the remaining boson contributions are written down as follows:
\al{
 \Pi_{\gamma\gamma}^{\text{T},s}|_{\text{{region II}}}(k^2) =
 \Pi_{\gamma\gamma}^{\text{T},s}|_{\text{odd}}(k^2)
 &={\alpha \over 4 \pi} \Bigg\{ -{2\over9} k^2 +\left( -\frac{1}{3} k^2 + \frac{4}{3} M_{W_s}^2 \right) B_{W_s}(k^2) - \frac{4}{3} M_{W_s}^2 B_{{W_s}}(0) \Bigg\},\\
 \Pi_{Z\gamma}^{\text{T},s}|_{\text{{region II}}}(k^2) =
 \Pi_{Z\gamma}^{\text{T},s}|_{\text{odd}}(k^2)
 &={\alpha \over 4 \pi} \sqbr{-\frac{c_W}{s_W}} \Bigg\{ -{2\over9} k^2 +\left( -\frac{1}{3} k^2 + \frac{4}{3} M_{W_s}^2 \right) B_{W_s}(k^2) - \frac{4}{3} M_{W_s}^2 B_{{W_s}}(0) \Bigg\}, \\
 \Pi_{ZZ}^{\text{T},s}|_{\text{{region II}}}(k^2) =
 \Pi_{ZZ}^{\text{T},s}|_{\text{odd}}(k^2)
 &={\alpha \over 4 \pi} \sqbr{\frac{c_W^2}{s_W^2}} \Bigg\{ -{2\over9} k^2 +\left( -\frac{1}{3} k^2 + \frac{4}{3} M_{W_s}^2 \right) B_{W_s}(k^2) - \frac{4}{3} M_{W_s}^2 B_{{W_s}}(0) \Bigg\}, \\
 \Pi_{WW}^{\text{T},s}|_{\text{{region II}}}(k^2) =
 \Pi_{WW}^{\text{T},s}|_{\text{odd}}(k^2)
 &= {\alpha \over 4\pi s_W^2} \Bigg\{ 
   - \frac{2}{9} k^2  
   - \frac{2}{3} M_{W_s}^2 B_{W_s}(0) - \frac{2}{3} M_{Z_s}^2 B_{Z_s}(0)
   \notag \\ 
 & \quad  +\left[ -\frac{1}{3} k^2 +\frac{2}{3} \left( M_{W_s}^2 + M_{Z_s}^2 \right) 
    \right] B_{W_s,Z_s}(k^2) \notag \\
 & \quad
   - \frac{1}{3} (M_{Z_s}^2 - M_{W_s}^2)^2 \delta B_{Z_s, W_s}(k^2) \Bigg \}.
}
The remaining part can be easily calculated by use of the following relations:
\al{
\Pi_{ab}^{\text{T},s}|_{\text{{region I}}}(k^2) &= 
\Pi_{ab}^{\text{T},s}(k^2) -
\Pi_{ab}^{\text{T},s}|_{\text{{region II}}}(k^2), \\
\Pi_{ab}^{\text{T},s}|_{\text{even}}(k^2) &= 
\Pi_{ab}^{\text{T},s}(k^2) -
\Pi_{ab}^{\text{T},s}|_{\text{odd}}(k^2),
}
where $ab$ represents the possible four combinations of gauge bosons.

We also derive the following relations for the mUED{:}
\al{
\Pi_{ab}^{\text{T},s}|_{\text{mUED}}(k^2) \simeq \Pi_{ab}^{\text{T},s}|_{\text{even}}(k^2)
}
based on the discussions in {Secs.} \ref{section:2} and \ref{section:ST}.


\section{{Summary of the bounds}
\label{Appendix:bounds}}
\begin{figure}
\begin{center}
   \includegraphics[viewport=0 0 955 325, width=32em]{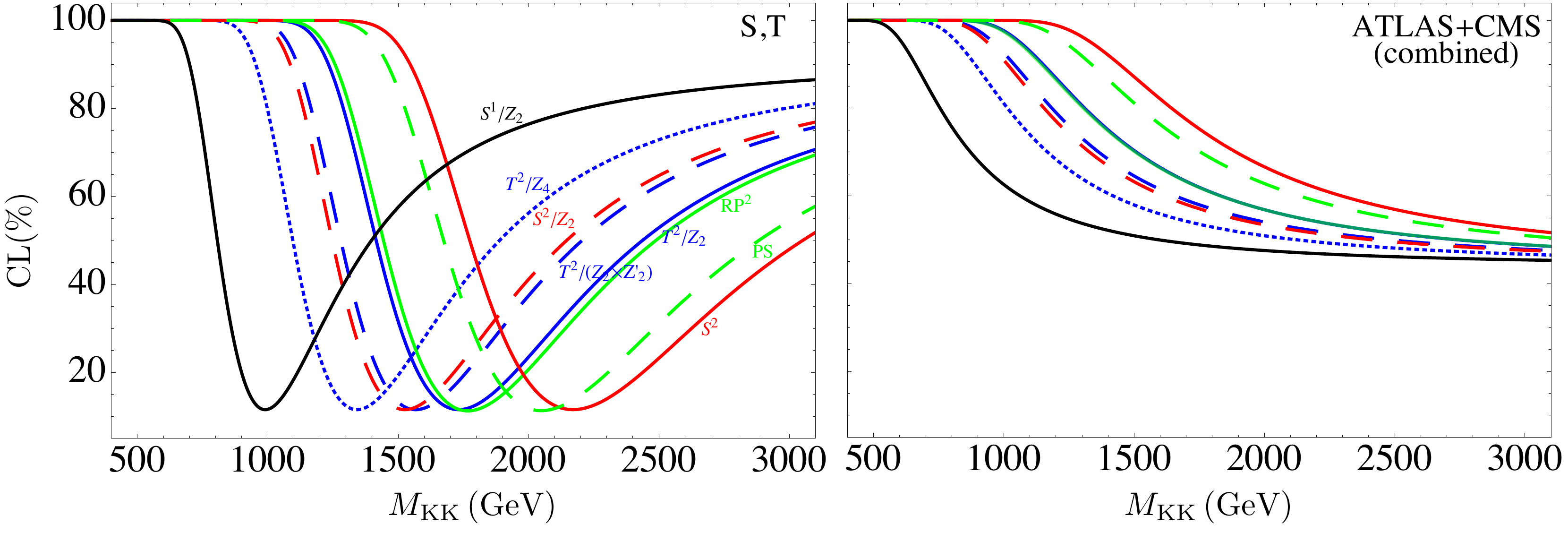}\\
   \includegraphics[viewport=0 0 955 325, width=32em]{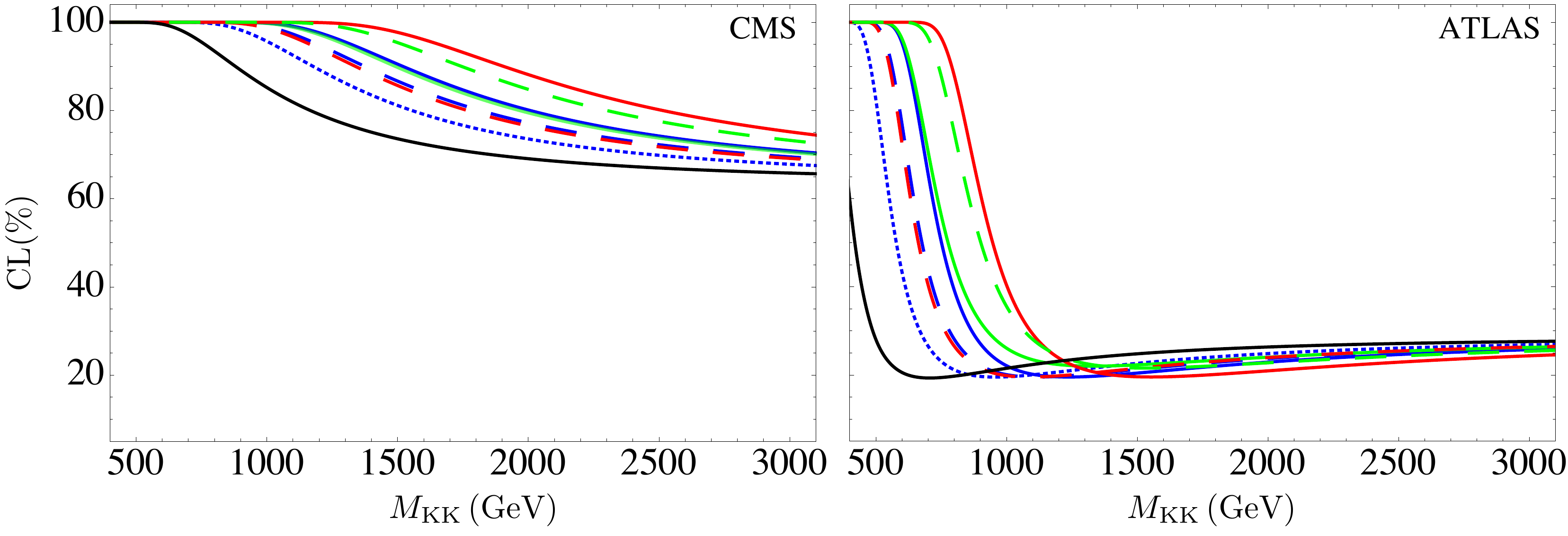}
\caption{
{
The exclusion {C.L.s} of all the UED models as functions of the KK scale $M_\text{KK}$ 
obtained from the experimental results of the Higgs searches at the LHC (ATLAS, CMS{,} and both of them{,} respectively) and {those} of $S,T$ parameters. 
Colors denote the same as in {Fig.}~\ref{fig:cutoff_upperbounds}. 
}
}
\label{Fig:UEDboundWide}
\end{center}
\end{figure}

Here{,} we summarize the bounds on the KK scale in UED models. 
Figure~\ref{Fig:UEDboundWide} shows the exclusion {C.L.s} as functions of the wide range of the KK scale. 


\bibliographystyle{TitleAndArxiv}
\bibliography{reference}

\end{document}